\documentclass{vldb}

\pdfoutput=1

\usepackage{graphicx}
\usepackage{amsmath}
\usepackage{verbatim}
\usepackage{amssymb}
\usepackage{subfigure}
\usepackage{times}
\usepackage{algorithm}
\usepackage{algpseudocode}
\usepackage{wrapfig, epsfig, epstopdf}
\usepackage{booktabs}
\usepackage{lipsum}
\usepackage{balance}

 \newcommand\blfootnote[1]{%
 	\begingroup
 	\renewcommand\thefootnote{}\footnote{#1}%
 	\addtocounter{footnote}{-1}%
 	\endgroup
 }

\newcommand{\+}{\mkern2mu}
\newcommand{\fall}{\+ \forall \+}

\newcommand{\trv}{tt}
\newcommand{\val}{val}
\newcommand{\ea}{ea}
\newcommand{\ld}{ld}
\newcommand{\pp}{\mathcal{P}}

\newcommand{\fwr}{\text{FWR}}
\newcommand{\fwri}{\text{FWR}$_i$}

\newcommand{\bwr}{\text{BWR}}
\newcommand{\bwrj}{\text{BWR}$_j$}

\newcommand{\gaps}{\texttt{gaps}}
\newcommand{\arcs}{\texttt{arcs}}

\newcommand{\forwest}{FWEST}
\newcommand{\eald}{EALD}

\newcommand{\static}{AOP}%{\textsc{Aop}}
\newcommand{\fastest}{Fastest}%{\textsc{Fastest}}
\newcommand{\twotd}{2TD-AOP}%{\textsc{2td-Aop}}
\newcommand{\algo}{\textsc{RecInsert}}
\newcommand{\prob}{2TD-AOP}

\newtheorem{definition}{Definition}
\newtheorem{lemma}{Lemma}

\title{Scenic Routes Now: Efficiently Solving the Time-Dependent Arc Orienteering Problem}

\begin{document}

\numberofauthors{7}

\author{
\alignauthor
Gregor Joss\'e$^\star$\\
	   \affaddr{LMU Munich}\\
       \affaddr{Oettingenstra\ss e 67, 80583 Munich, Germany}\\
       \email{josse@dbs.ifi.lmu.de}
\alignauthor
Ying Lu$^\star$\\
	   \affaddr{University of Southern California}\\
	   %\affaddr{Integrated Media Systems Center}\\
       %\affaddr{Powell Hall of Engineering, PHE 306}\\
       \affaddr{3737 Watt Way, Los Angeles, CA 90089-0272, USA}
       \email{ylu720@usc.edu}
\alignauthor %Dr.
Tobias Emrich\\
	   \affaddr{LMU Munich}\\
       \affaddr{Oettingenstra\ss e 67, 80583 Munich, Germany}\\
       \email{emrich@dbs.ifi.lmu.de}
\and
\alignauthor %Prof. 
Matthias Renz\\
	   \affaddr{George Mason University}\\
       \affaddr{4400 University Dr, Fairfax, VA 22030, USA}\\
       \email{mrenz@gmu.edu}
% 5th. author
\alignauthor %Prof.
Cyrus Shahabi\\
	   \affaddr{University of Southern California}\\
       %\affaddr{Integrated Media Systems Center}\\
       %\affaddr{Powell Hall of Engineering, PHE 306}\\
       \affaddr{3737 Watt Way, Los Angeles, CA 90089-0272, USA}
       \email{shahabi@usc.edu}
% 6th. author
\alignauthor %PD Dr.
Ugur Demiryurek\\
       \affaddr{University of Southern California}\\
       \affaddr{3737 Watt Way, Los Angeles, CA 90089-0272, USA}\\
       \email{demiryur@usc.edu}
}

\additionalauthors{Additional authors: Matthias Schubert (LMU Munich, Oettingenstra\ss e 67, 80583 Munich, Germany) {\texttt{schubert@dbs.ifi.lmu.de}})}

\maketitle

This paper extends the Arc Orienteering Problem (AOP) to large road networks with time-dependent travel times and time-dependent value gain, termed Twofold
Time-Dependent AOP or \prob{} for short. In its original definition, the NP-hard Orienteering Problem (OP) asks to find a path from a source to a destination
maximizing the accumulated value while not exceeding a cost budget. Variations of the OP and AOP have many practical applications such as mobile crowdsourcing
tasks (e.g., repairing and maintenance or dispatching field workers), diverse logistics problems (e.g., crowd control or controlling wildfires) as well as
several tourist guidance problems (e.g., generating trip recommendations or navigating through theme parks). In the proposed \prob{}, travel times and value
functions are assumed to be time-dependent. The dynamic values model, for instance, varying rewards in crowdsourcing tasks or varying urgency levels in damage
control tasks. We discuss this novel problem, prove the benefit of time-dependence empirically and present an efficient approximative solution, optimized for
fast response systems. Our approach is the first time-dependent variant of the AOP to be evaluated on a large scale, fine-grained, real-world road network. We
show that optimal solutions are infeasible and solutions to the static problem are often invalid. We propose an approximate dynamic programming solution which
produces valid paths and is orders of magnitude faster than any optimal solution.\blfootnote{$^{\star}$These authors contributed equally to this work.}

\section{Introduction}

Not long ago, finding the shortest or fastest path in road networks was the core challenge of route planning. Nowadays, challenges are manifold -- due to the
abundance of sensor data, due to the existence of increasingly complex traffic models and due to the demand for efficient yet convenient solutions. Modern
navigation systems often take multiple, sometimes time-dependent, cost criteria into account and solve complex queries in order to increase driver convenience
and traffic efficiency. In view of this progress, we propose a novel query with multiple applications. As an extension to the family of Orienteering Problems,
we present the Twofold Time-Dependent Arc Orienteering Problem (\prob{}).

In its original formulation \cite{OriginalPaper-87}, the Orienteering Problem (OP) requires to find a path from a given source to a given destination 
abiding by a given cost budget but maximizing the value collected along the way. In most cases, the cost function corresponds to the travel time within
the given network. The problem may for instance occur in the field of spatial crowdsourcing
\cite{CSMTA-GeocrowdEnablingQueryAnsweringWithSpatialCrowdsourcing12SHAHABI,CS-Chen-VLDB-15,CS-Shahabi-ACM-15} where individuals complete advertised tasks in
order to collect rewards. A recent popular example of this is the augmented reality game Pok\'emon Go. A worker who is willing to complete crowdsourced tasks
will want to maximize the collected rewards within his chosen time frame. The OP combines the NP-hard Knapsack and Traveling Salesman Problems, hence is NP-hard
itself. Numerous variations of the problem have been proposed, for instance, extending it to multiple workers (Team Orienteering Problem)
\cite{OP-Survey-Vansteenwegen-16} or restricting the collectable values to certain time windows ((Team) Orienteering Problem with Time Windows)
\cite{TTDP-Survey-Gavalas-14}. Another variation is the Arc Orienteering Problem (AOP) \cite{AOP-Survey-ArchettiSperanza-13}.
In the OP, the value is associated with the vertices of the graph, whereas in the AOP, it is associated with the arcs. Thus, the OP intuitively corresponds to logistic or
crowdsourcing tasks, where the value is collected at particular points. The AOP, on the other hand, corresponds to tasks where the value is collected ``along
the way''. Examples of such tasks are the routes of firefighting planes or the planning of scenic bike trips
\cite{AOP-GRASP-PlanningOfCycleTripsInEastFlanders-VANSTEENWEGEN-OMEGA10}.

Until recently, the AOP has only been answered for networks with static travel times. However, assuming static costs is not reasonable for most real-world
scenarios. For road networks, the significance of
time-dependence was empirically proven in \cite{TD-CaseForTimeDepRoadNetworks-FoundationPaper-DEMIR-SHAHABI-GIS10} and substantiated by the incorporation into
many navigation services such as Waze\footnote{waze.com}. In a recent study, Verbeeck et. al 
\cite{TDOP-YING-AntColony-DriveSpeedModel-FastSolutionToTDOP-OR14-VANSTEENWEGEN} first introduced time-dependent travel times but the values have been
assumed to be static. In this paper, we propose to incorporate time-dependent values in addition to time-dependent travel times, introducing the \prob{}.
As will be shown, results benefit greatly from time-dependent values which model the varying benefit for different times. Without loss of
generality, we restrict ourselves to the application of scenic route planning. In this use case, a view of the sea might only be worthwhile during day, whereas
a view of the skyline might be particularly worthwhile at night. Thus, the value of a coastline road is high during the day but low at night, while a road
passing through downtown may have the opposite values. Other use cases include the aforementioned firefighting planes, collecting street view data or gathering
Pok\'emon Go items.
 
\begin{figure}[t]
\centering
        \subfigure[Fastest path]{
            \includegraphics[width=0.45\columnwidth]{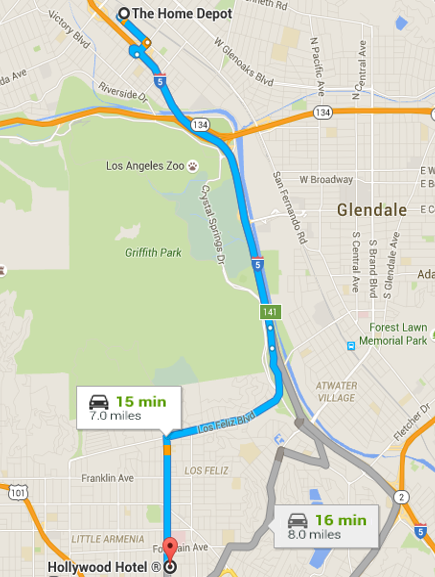}
            \label{fig:motiv-fastest}
            }
        \hfill
        \subfigure[Static scenic path]{
            \includegraphics[width=0.45\columnwidth]{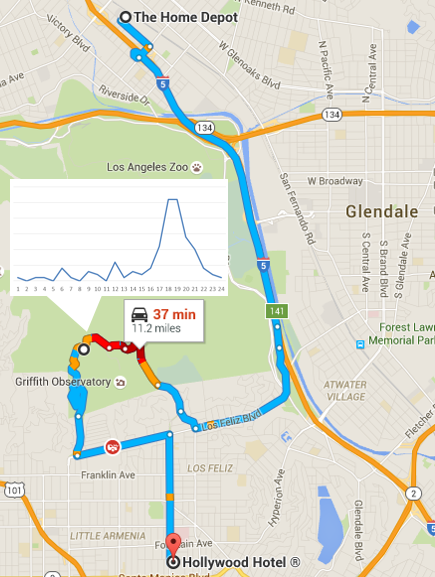}
            \label{fig:motiv-static}
            }\\
       \subfigure[Time-dependent  scenic path (11:00 am)]{
            \includegraphics[width=0.45\columnwidth]{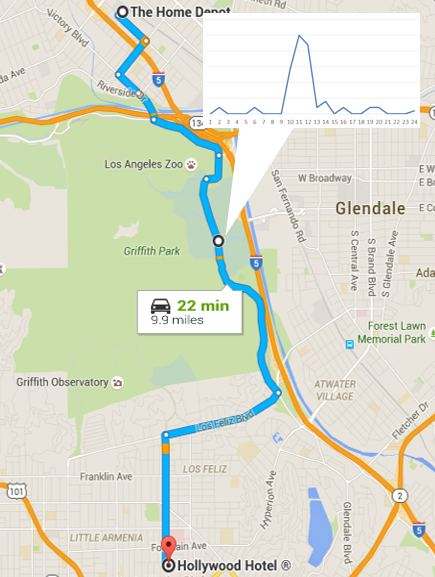}
            \label{fig:motiv-td-1}
            }     
        \hfill
        \subfigure[Time-dependent scenic path (06:00 pm)]{
            \includegraphics[width=0.45\columnwidth]{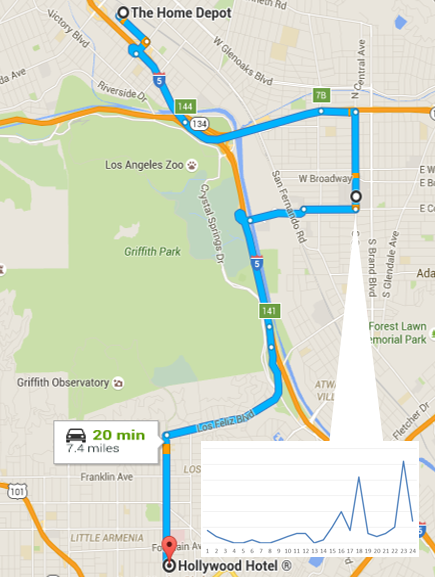}
            \label{fig:motiv-td-2}
            }
        \caption{Exemplary paths with hourly value distribution for particular areas along the path.}
            \label{fig:motiv}
\end{figure}

Consider the example given in Figure~\ref{fig:motiv} which shows different result paths for a scenic path query with the same source and destination.
Figure~\ref{fig:motiv-fastest} shows the time-dependent fastest path according to our algorithm for two departure times.  While the path is the same at 11
am and 6 pm, the travel times are not the same (15 and 20 minutes, respectively). In this example, the fastest path does not pass through particularly scenic
areas, according to the crowdsourced data set. The other paths pass through such areas, and for particular scenic parts of the respective paths, the hourly
value distributions are highlighted. At first glance it is evident that the value functions show great variance over the course of a day. If not allowing for
time-dependent values, these functions are commonly averaged, evening out the significant peaks of the distribution.

Suppose a driver is willing to spend 25 minutes for his trip from Hollywood Hotel to the Home Depot north of Griffith Park. Employing an algorithm for the
AOP in static networks\cite{AOP-Ying-Shahabi-ACMGIS-15}, we obtain the \emph{static scenic path} displayed in Figure~\ref{fig:motiv-static}. The path was
computed in a static network derived from our time-dependent network. The static travel time along an arc is set as the average of all travel times along this
arc in the time-dependent network. The static scenicness values are derived analogously. The generated result path has a static travel time of 22 minutes which
abides by the budget. However, the actual travel time might of course deviate in either direction. For a desired departure at 11 am, the time-dependent
travel time coincides with the static time of 22 minutes, but for a desired departure at 6 pm, the time-dependent path takes 37 minutes. Thus, at 6 pm
the result path clearly exceeds the budget and is therefore invalid. In a similar way, the averaged scenicness values can deviate from the time-dependent
ones. While the static scenic path at 11 am does not exceed the budget, it is not optimal either. Griffith Observatory, which it passes along the way, is a
popular sight but particularly scenic during sunset or in the evening. Instead, the best solution at 11 am is displayed in Figure~\ref{fig:motiv-td-1}. This
\emph{time-dependent scenic path} takes 20 minutes and passes a golf course, offering particularly nice views at daytime. The best solution at 6 pm is shown
in Figure~\ref{fig:motiv-td-2}. This path takes 22 minutes and passes through central Glendale, a bustling area in the evening and at night. 

Taking time-dependence into account when computing AOP paths is crucial. Not only result quality but also result validity depends on it. Our experiments show
that every second static solution in a moderately complex settings is invalid. Users benefit considerably from time-dependent travel times as every \prob{}
path is valid. Furthermore, results are improved significantly when the values reflect time-dependence as well. Our experiments show that \prob{} paths generate
three to four times more value than static solutions. We therefore propose to employ value distribution models whenever possible.

Due to their NP-hardness, static variations of the OP and AOP are usually solved employing metaheuristics. Allowing for time-dependent travel times adds further
complexity to the task. Slight modifications of valid solutions may invalidate them. For instance, when taking a minor detour at the beginning of the path,
travel times for the rest of the path may change and exceed the budget. Additionally incorporating time-dependent values multiplies this effect. Taking a minor
detour at the beginning of a path may result in a change of value and travel time. We propose to solve the \prob{} with a novel dynamic programming approach
which makes use of pruning techniques to limit the candidate set and reduce intermediate path computations.

To the best of our knowledge, no previous work has allowed for both travel times and values to be time-dependent. In addition to handling the increased
complexity, we require fast response times.	We focus on the practicality of the solution where fast response times are crucial and large road network are
standard, for instance, in order to meet the requirements of a mobile application. In conclusion, the contributions of this paper are as follows:

\begin{enumerate}
  \item We motivate and introduce the \prob{}, as a new graph-based routing problem and a Mixed Integer Programming formulation.
  \item We present an efficient approximate solution to the \prob{}.
  \item We present experiments on a large-scale real-world road network. Our evaluation is the first of this dimension for any time-dependent AOP. We show the
  value of twofold time-dependence empirically and compare \prob{} solutions to static solutions as well as to optimal results.
\end{enumerate}

The remainder of this paper is organized as follows. In Section~\ref{sec:related} we give an overview of research relevant to our work. In
Section~\ref{sec:problem-definition} the \prob{} is formally defined and formulated as an Mixed Integer Programming. In Section~\ref{sec:solution} we lay the
theoretical foundation for our algorithm which is described in Sections~\ref{sec:algorithm-1} and \ref{sec:algorithm-2}. Subsequently, our algorithm is
evaluated in Section~\ref{sec:exp}. A conclusion is provided in Section~\ref{sec:conclusion}.

\section{Related Work}
\label{sec:related}

We divide research relevant to this work into three groups. First, we review research on how to attain profit values from data in general and crowdsourced
content in particular. Next, we briefly introduce related research from the database community such as the Trip Planning Query (TPQ). These problems have a
similar field of application but are not congruent to the problem discussed in this paper. Finally, we present research on the family of OPs which
mostly stems from the field of Operations Research. We give an overview of solutions to the static and time-dependent OP and AOP. 

\subsection{Attaining Value Functions}
An important aspect of the OP/AOP is how to attain the value functions, in particular when aiming to solve a time-dependent instance. Most works from the field
of Operations Research rely on specific bench mark data sets, some synthetic, others real. The authors of
\cite{AOP-GRASP-PlanningOfCycleTripsInEastFlanders-VANSTEENWEGEN-OMEGA10}, for instance, evaluate their approach to cycle trip planning on real-world tourist
data from Flanders, Belgium. The authors of \cite{TDOP-MathematicalModelMetaheuristicsTDOP-Gunawan2014} use data from Chinese theme parks.
Relying on specifically recorded data like the above has two major drawbacks. First, expert knowledge or an official mandate may be required to
attain the data. Second, the data may not be available globally. As this approach usually lacks scalability, there exist advances to mine this knowledge from
freely available data.

Crowdsourced data is a rich source of information. It is for instance used for general POI recommendation
\cite{TDOP-YING-PredOfCIKM14-ExploitingLargeScaleCheckInDataToRecommendTimeSensitiveRoutes-UrbComp12}, for categorization
\cite{TDVCOP-TripPlannerPersTripPlanLeveragingHeterogenCrowdsourcedDigiFootprints-TITS14}, for popularity estimation
\cite{POI-Preferences-Categories-GenMaxCover-PlanningTouristicToursWithTripBuilder-CIKM13}, for determining opening hours or visiting times
\cite{TDOP-YING-Gowalla-CheckIns-MiningPlanningTimeAwareRoutesfromCheckInData-CIKM14}, for determining duration of stay
\cite{POI-VisTimeFromFlickrPics-AutomaticConstructionOfTravelItinerariesFromSocialBreadcrumbs-YAHOO10} or for determining spatial connectivity
\cite{SSTD15-Knowledge}. The most common data sources for these purposes are Foursquare, Yelp,
Twitter, Flickr\footnote{foursquare.com, yelp.com, twitter.com, flickr.com} or the inoperative but oft-cited Gowalla. When interested
in how to attain value functions from the raw data, we refer the reader to the respective works. In the context of this paper, we assume the values to be given. For our experimental
evaluation, we rely on Flickr data, details are given in Section~\ref{sec:exp}.

\subsection{TPQ and Related Queries}

The most related query in the database community is the TPQ \cite{TPQ} (alternatively, Route Planning Query \cite{MRPSRQ} or Route
Search Query \cite{LevKanSafSag09}). In this problem setting, the user specifies a different POI categories, e.g., ``ATM'', ``restaurant'', ``florist'',
``cinema''. The result of a TPQ is the fastest path from a given source to a given destination visiting exactly one instance of each resource. The TPQ is
NP-hard because in the case that each category occurs exactly once, the TPQ degenerates to the Traveling Salesman Problem. There exist several variations, for
instance introducing order constraints \cite{OSR,InteractiveRouteSearchPresenceOrderConstraints} or modeling the fulfillment at a location probabilistically
or time-dependent \cite{FollowUpToKRoute,EDBT15-ResourceRoute,DelEfsVer11}. Solutions to this problem include full enumeration \cite{DelEfsVer11}, heuristics
\cite{MRPSRQ,OSR,DelEfsVer11}, or backtracking and branch-and-bound approaches \cite{EDBT15-ResourceRoute}. Partially time-dependent variations of these queries
exist, however, most of the research pertains to static networks. Most importantly, albeit related, the problem settings of the OP/AOP and the TPQ are not
congruent.

\subsection{OP and AOP}

Blueprint for the OP was the family of orienteering sports where efficient navigation from checkpoint to checkpoint is the goal \cite{OriginalPaper-87}.
Alternative names for the OP include the Selective Traveling Salesman Problem \cite{OP-STSP} or the Traveling Salesman Problem with Profits 
\cite{OP-TSPwithProfits}. The OP is executed on a graph whose arc weights represent traversal costs and whose vertices may yield certain value. Given a source, a
destination and a cost budget, the output of an OP is the path with maximum value among all paths not exceeding the budget. In the variation referred to as
AOP or Privatized Rural Postman Problem \cite{AOP-PrivatizedRuralPostmanProblems}, the profit is not assigned to vertices but to arcs, while the rest is as
before. Extensive surveys on these problems and their numerous variants are given in
\cite{OP-Survey-Vansteenwegen-11,OP-Survey-Vansteenwegen-16,AOP-Survey-ArchettiSperanza-13} (or from the perspective of the related Tourist Trip Design Problem
in \cite{TTDP-Survey-Gavalas-14}).

Although NP-hard \cite{OriginalPaper-87}, exact solutions to the OP exist \cite{OP-STSP,OP-Exact-Ramesh-91,OP-TSPwithProfits}. Most solutions, however, are
approximative. While in older research specific heuristic approaches were proposed \cite{OP-Heuristics-84,OP-Heuristic-Golden-88,OP-Heuristic-Chao-96}, the
trend shifted towards algorithms following particular meta-heuristics such as Tabu Search \cite{OP-TabuSearch-Gendreau-98}, Genetic Algorithms
\cite{OP-Genetic-Tasgitiren-01} or Ant Colony Algorithms \cite{OP-AntColony-Liang-02}. This holds similarly for the AOP where exact algorithms
\cite{AOP-Exact,AOP-ILS-ExtensionOfAOPAndApplicationCycleTripPlanning-VANSTEENWEGEN-TR14} are rare but available. Generating approximative solutions is the
standard approach \cite{AOP-AppoximationAlgorithms-Gavalas-14}, most effectively following the meta-heuristics Iterative Local Search (ILS)
\cite{AOP-ILS-ExtensionOfAOPAndApplicationCycleTripPlanning-VANSTEENWEGEN-TR14} and Greedy Randomized Adaptive Search Procedure (GRASP)
\cite{AOP-GRASP-PlanningOfCycleTripsInEastFlanders-VANSTEENWEGEN-OMEGA10,AOP-Ying-Shahabi-ACMGIS-15}. Recently, ILS and GRASP have been improved to solve the
AOP on large scale real-world road networks in near-interactive time \cite{AOP-Ying-Shahabi-ACMGIS-15}.

Introduced in \cite{TDOP-NPhardness-FirstApprox-Fomin-02}, the OP with time-dependent (TD-OP) costs has gained attention recently
\cite{TDOP-MathematicalModelMetaheuristicsTDOP-Gunawan2014,TDTOPTW-Garcia-First-09,TDTOPTW-EfficientHeuristicsForTDTOPTW-13,TDOP-YING-AntColony-DriveSpeedModel-FastSolutionToTDOP-OR14-VANSTEENWEGEN,TDOP-ResearchOnTOPwithDynamicTT-Li-12}.
Typical applications for the TD-OP include electronic tourist guides \cite{TDTOPTW-Garcia-First-09,TDTOPTW-POI-PublicTransport-PersonalizedTouristRouteGen-10}
or routing in theme parks \cite{TDOP-MathematicalModelMetaheuristicsTDOP-Gunawan2014}. The time-dependent AOP (TD-AOP) as presented in
\cite{TDOP-YING-AntColony-DriveSpeedModel-FastSolutionToTDOP-OR14-VANSTEENWEGEN} is not directly linked to an application but a time-dependent extension of
the authors' cycle trip planning application \cite{AOP-ILS-ExtensionOfAOPAndApplicationCycleTripPlanning-VANSTEENWEGEN-TR14} is imaginable.

The above works are all evaluated on small graphs (with the exception of \cite{AOP-Ying-Shahabi-ACMGIS-15}) with no more than a couple of hundred vertices. 
Most arcs in these graphs connect two non-zero value vertices. For instance, in the theme park
data set of \cite{TDOP-MathematicalModelMetaheuristicsTDOP-Gunawan2014}, vertices correspond to attractions and the arcs are fastest paths between them. The
travel times and the values are derived from user data of the theme parks and their attractions. Similarly, \cite{TDTOPTW-EfficientHeuristicsForTDTOPTW-13} and
the demonstration paper \cite{TDTOPTW-POI-PublicTransport-PersonalizedTouristRouteGen-10} are evaluated on data sets consisting of Points of Interest (POIs)
and fastest paths connecting these POIs (by walking or by public transportation). When considering real-world road networks, this is not feasible. A
moderate-sized network usually contains over one million vertices and two to three million arcs. In a time-dependent network, there usually exist numerous
fastest paths from a given vertex to another, depending on the departure time. Hence, introducing shortcuts for precomputed fastest paths as in the
approaches above is not possible.  This is particularly the case, if the travel times are not statistically inferred but real-time, i.e., depending on live
information, as is the case in the network of our evaluation.

The most closely related problem to the \prob{} is the TD-AOP with time-dependent costs and static value. Verbeeck et. al recently proposed to solve this
problem with an Ant Colony System algorithm \cite{TDOP-YING-AntColony-DriveSpeedModel-FastSolutionToTDOP-OR14-VANSTEENWEGEN}. Evaluated on specific benchmark
instances with at most 100 vertices, the proposed approach finds near-optimal to optimal solutions within an average runtime of one second. The
complex solver  has excessive time requirements (``often more than 100 hours'' \cite{TDOP-YING-AntColony-DriveSpeedModel-FastSolutionToTDOP-OR14-VANSTEENWEGEN})
to generate optimal solutions on these benchmark instances of the TD-AOP. In comparison, we increase complexity of the problem in two ways. First, by including
time-dependent value functions \prob{} adds another dimension of time-dependence to the problem. Second, we aim to solve the problem on large-scale real-world road
networks with hundred thousands of vertices. Established solutions to related problems like the TD-AOP or the TD-OP rely on metaheuristics such as the Ant
Colony System \cite{TDOP-YING-AntColony-DriveSpeedModel-FastSolutionToTDOP-OR14-VANSTEENWEGEN}, Iterative Local Search
\cite{AOP-ILS-ExtensionOfAOPAndApplicationCycleTripPlanning-VANSTEENWEGEN-TR14} or Greedy Randomized Adaptive Search Procedure
\cite{AOP-GRASP-PlanningOfCycleTripsInEastFlanders-VANSTEENWEGEN-OMEGA10,AOP-Ying-Shahabi-ACMGIS-15}). Instead, we propose a heuristic dynamic programming
approach.

\section{Problem Definition}
\label{sec:problem-definition}

We model the road network as a graph $G = (V, A, \val, \trv)$, where $V$ denotes the set of vertices (or nodes), $A\subseteq V\times V$ denotes the set of
directed arcs (or edges), $\val:V\times V\times \mathbb{R}_0^+\rightarrow \mathbb{R}_0^+$ and $\trv$ (with the same domains) denote the
non-negative functions mapping each arc onto its respective value and the travel time, respectively
\[\val_{i,j}(t) := \val(v_i, v_j,t) \geq 0, \quad \trv_{i,j}(t) := \trv(v_i,v_j,t) \geq 0\]
For the representation of the time-dependent travel time along an arc, we use a piecewise constant function with equal step width $\tau$. In the
following, we refer to the intervals where travel time is constant as time windows and denote the $k$-th time window by $\tau_k$, where $0\leq k\leq K$ for
some maximum time window $K$.
% For the road network in our experiments, for instance, $\tau = 5$ (?). When issuing a query, departure times $t_0$ must be specified. W.l.o.g.\ we assume $t_0 =
% 0$, i.e., the time domain begins at departure time. Note that in this case the first time window $\tau_0$ may be shorter than $\tau$. Similarly, we assume a
% piecewise constant value function with time windows $\tau'_k$ of constant length $\tau'$.

% For each query, we denote the source vertex by $v_s$ and the destination vertex by
% $v_N$. Note that these indexes are query-dependent and do not refer to fixed vertices. 

A consecutive set of arcs which visits no arc twice is called a path. A path from source to destination departing at $t_0$ is denoted by
$p = ((v_{p_0},v_{p_1}),(v_{p_1},v_{p_2}),\dots,(v_{p_{N-1}},v_{p_N}), t_0)$
where $v_{p_0}$ is the source vertex and $v_{p_N}$ the destination vertex. In the remainder of this paper, we simplify notation for the sake of
clarity. We omit the double indexes and denote a path $p$ by \[p=((v_0,v_1),(v_1,v_2),\dots,(v_{N-1},v_N), t_0)\]
For each query, we denote source and destination by $v_0$ and $v_N$, respectively.
The travel time of a path $p$ is defined as $\trv(p)
= \sum_{i=0,\dots,N-1} \trv_{i,i+1}(t_i)$, where $t_i$ denotes the time of arrival at (and departure from) vertex $v_i$. $t_i$ is dependent on the travel times
along the preceding arcs and is defined iteratively:
\[t_i := \sum_{j=0}^{i-1}\trv_{j,j+1}(t_j)\]
The value of a path $p$ is defined as:
\[\val(p) = \sum_{i=0,\dots,N-1} \val_{i,i+1}(t_i)\]
For a given source $v_0$, destination $v_N$, and departure time $t_0$, we denote the set of all paths from $v_0$ to $v_N$ starting at $t_0$ by $\pp_{0,N}(t_0)$.

The input for the \prob{} is a graph $G$, a source vertex $v_0$, a destination vertex $v_N$, a departure time $t_0$, and a time budget $b$. When speaking of a
query, we mean the set of these inputs. The optimal path to a \prob{} query is the path maximizing the collected value among all feasible paths. Both
notions are defined in the following.

\begin{definition}[feasible path]
Given a query, we call any path $p$ departing from $v_0$ at $t_0$ and arriving at $v_N$ no later than $t_0 + b$ (i.e., $\trv(p) \leq b$) a \emph{feasible path}. 
\end{definition}

\begin{definition}[optimal path]
Given a query, among all feasible paths, we call one with maximum value an \emph{optimal path}.
\end{definition}

Extending the work in \cite{OP-Survey-Vansteenwegen-11,TDOP-YING-AntColony-DriveSpeedModel-FastSolutionToTDOP-OR14-VANSTEENWEGEN}, we give a mixed
integer programming (MIP) formulation of the problem. We introduce the following two decision variables: $p_{i,j,k}$ is a binary decision variable which equals
$1$ if the result path enters the arc $(v_i,v_j)$ in time window $\tau_k$ and $0$ if the arc is not traversed at all or entered in a different time window.
$t_{i,j,k}$ is a continuous decision variable containing the time of departure at $v_j$ when entering the arc $(v_i, v_j)$ in time window $\tau_k$. Again,
it is $0$ if the arc is not traversed at all or entered in a different time window.

\vspace{-1eM}
\begin{subequations}
\begin{align}\label{mip:value}
\max \sum_{i\neq j} \sum_{k=0}^K p_{i,j,k} \cdot \val_{i,j}(\tau_k)
\\
\label{mip:budget}
\text{subject to: }\quad\quad\quad\quad\quad\sum_{i\neq j} \sum_{k=0}^K p_{i,j,k} \cdot \trv_{i,j}(\tau_k) \leq b & 
\\
\label{mip:once}
\fall 1 \leq l \leq N-1 \notag:
\\\sum_{i=0}^{N-1}\sum_{k=0}^K p_{i,l,k} = \sum_{j=1}^{N}\sum_{k=0}^K p_{l,j,k} \leq 1
\\
\label{mip:sourcedest}
\sum_{j=1}^N p_{0,j,0} = \sum_{i=0}^N\sum_{k=0}^K p_{i,N,k} = 1 & 
\\
\label{mip:departure}
\fall 1 \leq l \leq N-1 \notag:
\\\sum_{i=0}^{N-1} \sum_{k=0}^K t_{i,l,k} + \trv_{i,l}(\tau_k) = \sum_{j=1}^{N}\sum_{k=0}^K
t_{l,j,k}
\\
\label{mip:contraints1}
p_{i,j,k} \in \{0,1\}, \ \fall i,j,k
\\
\label{mip:contraints2}
0 \leq t_{i,j,k} \leq \tau_K, \ \fall i,j,k 
\\
\label{mip:contraints3}
t_{0,j,0} = t_0, \ \fall j
\end{align}
\end{subequations}

Equation~\ref{mip:value} gives the objective function while Equation~\ref{mip:budget} gives the budget constraint. Note that due to the time-dependent nature of
the problem, all time windows have to be taken into account when summing the collected value and cost. Remember, however, that $p_{i,j,k}$ is a binary decision
variable which for each arc can at most once be equal to $1$, as each vertex (and thereby also each arc) is at most visited once. This is expressed in
Constraint~\ref{mip:once}. Constraint~\ref{mip:sourcedest} ensures that the result path starts at the source and ends at the destination. Next,
Constraint~\ref{mip:departure} ensures that the departure time at each arc is the departure time at its predecessor plus the travel time within the
corresponding time window. Note that this condition implicitly forbids ``waiting'' at vertices. Finally,
Constraints~\ref{mip:contraints1}~and~\ref{mip:contraints2} determine the domain of the decision variables and Constraint~\ref{mip:contraints3} ensures that
the path departs from the source at $t_0$.

\section{Solving the \prob{}}
\label{sec:solution}

In this section, we present terminology and insights essential to our solution of the \prob{}. 

\begin{definition}[earliest arrival]
Given a graph $G$, a source $v_0$, a vertex $v_i$, and a departure time $t_0$, we define the \emph{earliest arrival (time)} for reaching $v_i$ from $v_0$
when departing at $t_0$, denoted by $\ea_{0,i}$ ($\ea_i$ for short), as follows:
\[ \ea_i := t_0 + \min_{p\in\pp_{0,i}(t_0)} \trv(p) \]
\end{definition}

\begin{definition}[latest departure]
Given a graph $G$, a destination $v_N$, a vertex $v_i$, and a latest arrival time $t_0+b$, we define the \emph{latest departure (time)} for reaching $v_N$ from
$v_0$ no later than $t_0+b$, denoted by $\ld_{i,N}(t_0+b)$ ($\ld_i$ for short), as follows:
\[\ld_i := t_0 + b - \min_{p\in\pp_{i,N}(-\infty)}\trv(p)\]
i.e., the latest departure time among all feasible paths w.r.t. $v_i$, $v_N$, and $t$.
\end{definition}

\begin{definition}[forward-reachable]
Given a graph $G$, a source $v_0$, a departure time $t_0$, and a time budget $b$, we refer to the vertices reachable within the budget $b$ from $v_0$ (when
departing at $t_0$) as \emph{forward-reachable} vertices, denoted by $\fwr_0(b)$ ($\fwr{}$ for short if $v_0$ and $b$ are clear from context).
\[\fwr_0(b) := \{v_i\in V \ \mid \ \ea_i \leq t_0 + b\}\]
\end{definition}

\begin{definition}[backward-reachable]
Given a graph $G$, a destination $v_N$, a departure time $t_0$, and a time budget $b$, we refer to the vertices from which the destination can be reached within
the budget $b$ when departing no earlier than $t_0$ as \emph{backward-reachable} vertices, denoted by $\bwr_N(b)$ ($\bwr{}$ for short).
\[\bwr_N(b) := \{v_i\in V \ \mid \ \ld_i \geq t_0\}\]
\end{definition}

\begin{figure}[t]
	\centering
	\subfigure[$\fwr_0(b)$ and $\bwr_N(b)$ and their intersection]{
		\includegraphics[width=0.46\columnwidth]{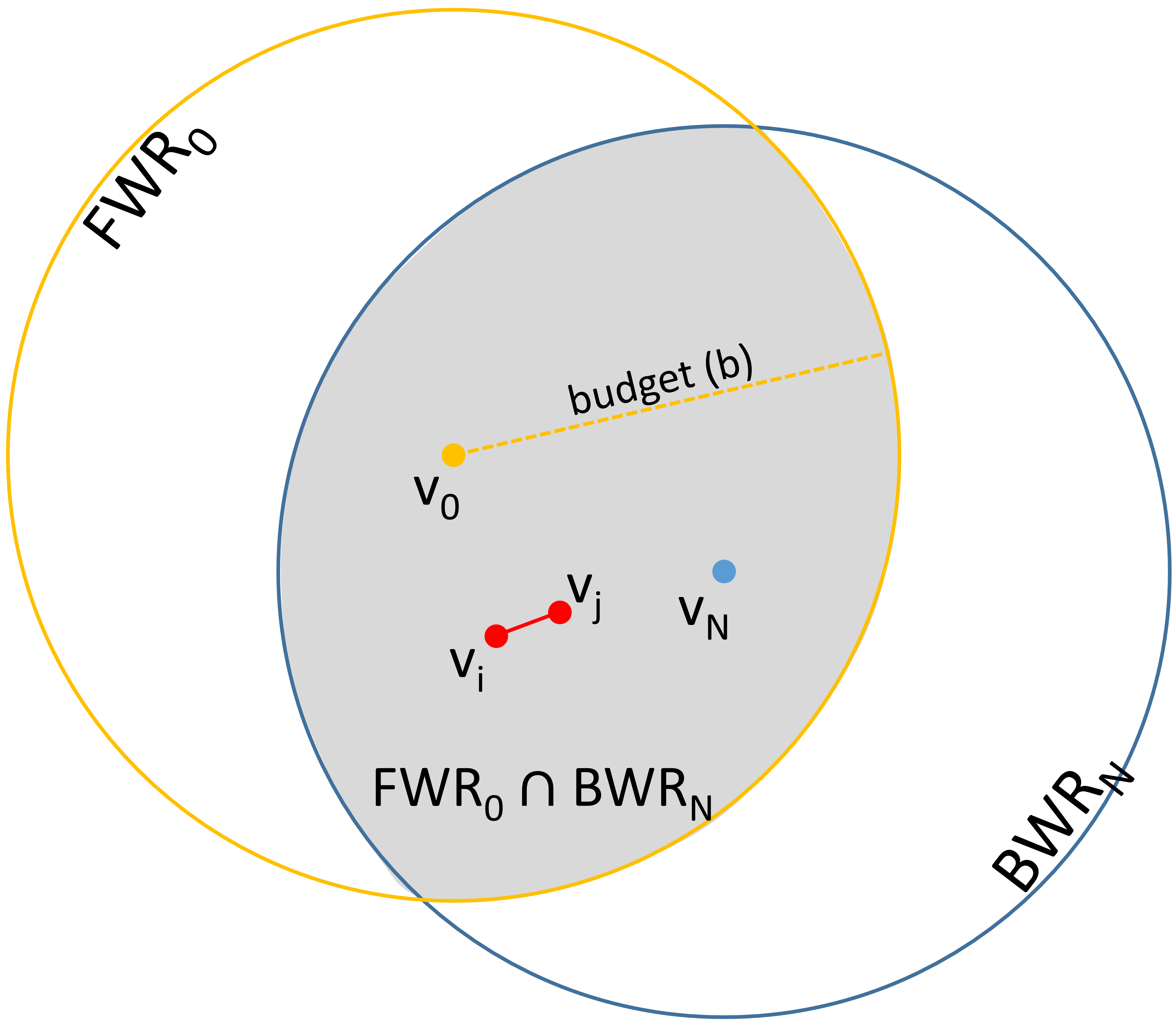}
		\label{fig:balls-1}
	}
	\hfill
	\subfigure[Recursive inclusion of the reachability regions and their intersection]{
		\includegraphics[width=0.47\columnwidth]{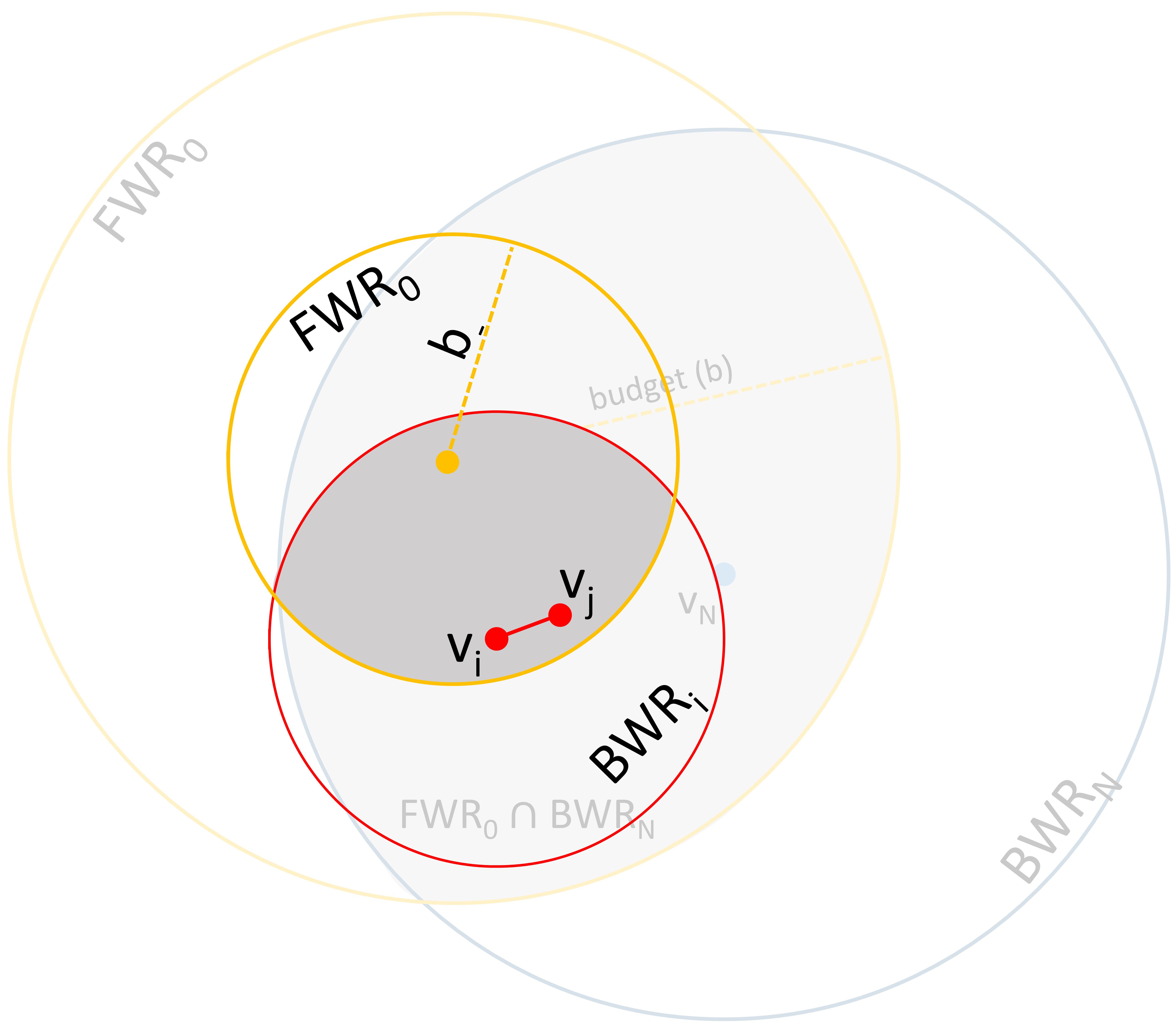}
		\label{fig:balls-2}
	}
	\caption{Reachability regions as computed by \fwr{} and \bwr{}.}
	\label{fig:balls}
\end{figure}

Intuitively, all forward-reachable vertices lie in a quasi-circular region around the source. This region is similar to the circular
expansion of a Dijkstra-search. The difference is that in this case the cost criterion is time-dependent travel time. Analogously,
all backward-reachable vertices lie in a quasi-circular region around the destination. This graphical intuition is visualized in
Figure~\ref{fig:balls-1}. If the destination is not forward-reachable from the source (and/or vice versa), then
the query has no answer. 

Our solution to the \prob{} is based on an extension of this observation: For an arc $(v_m,v_n)$ to be part of a
feasible path, it is a necessary condition that $v_m$ is forward-reachable from the source $v_0$ and $v_n$ is backward-reachable from the
destination $v_N$. Hence, if an arc is not contained in the intersection of the forward-reachability region and the backward-reachability region,
it cannot be part of a feasible path. This property will be proved later on. In our solution to the \prob{}, arcs which increase the
value are recursively inserted until the budget is exhausted. With every insertion of an arc, the overall travel time of the path increases. Hence, the
reachability regions iteratively contract until no more arcs can be inserted. Before we detail this approach in the next section, let us explain the above
definitions and their computation with an example.

\begin{figure}[t]
\centering
        \subfigure[Example graph]{
            \includegraphics[width=0.46\columnwidth]{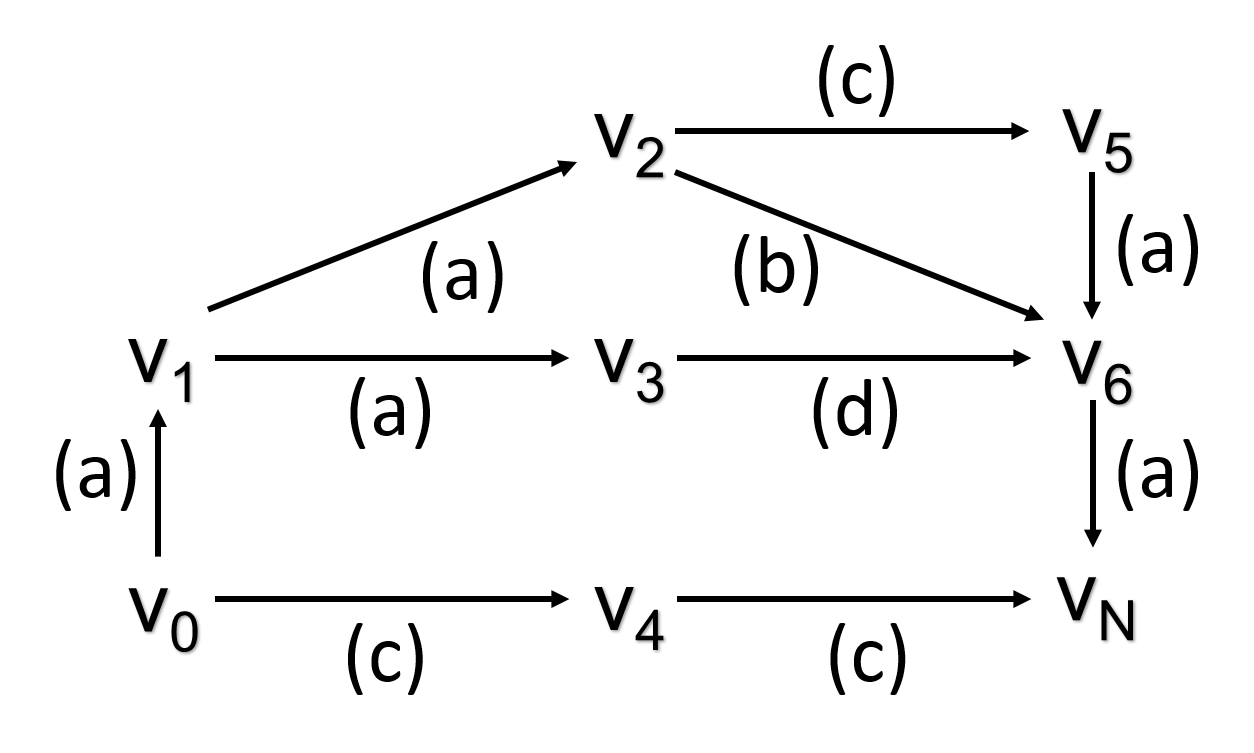}
            \label{fig:ex-0}
            }
        \hfill
        \subfigure[Example travel times functions]{
            \includegraphics[width=0.47\columnwidth]{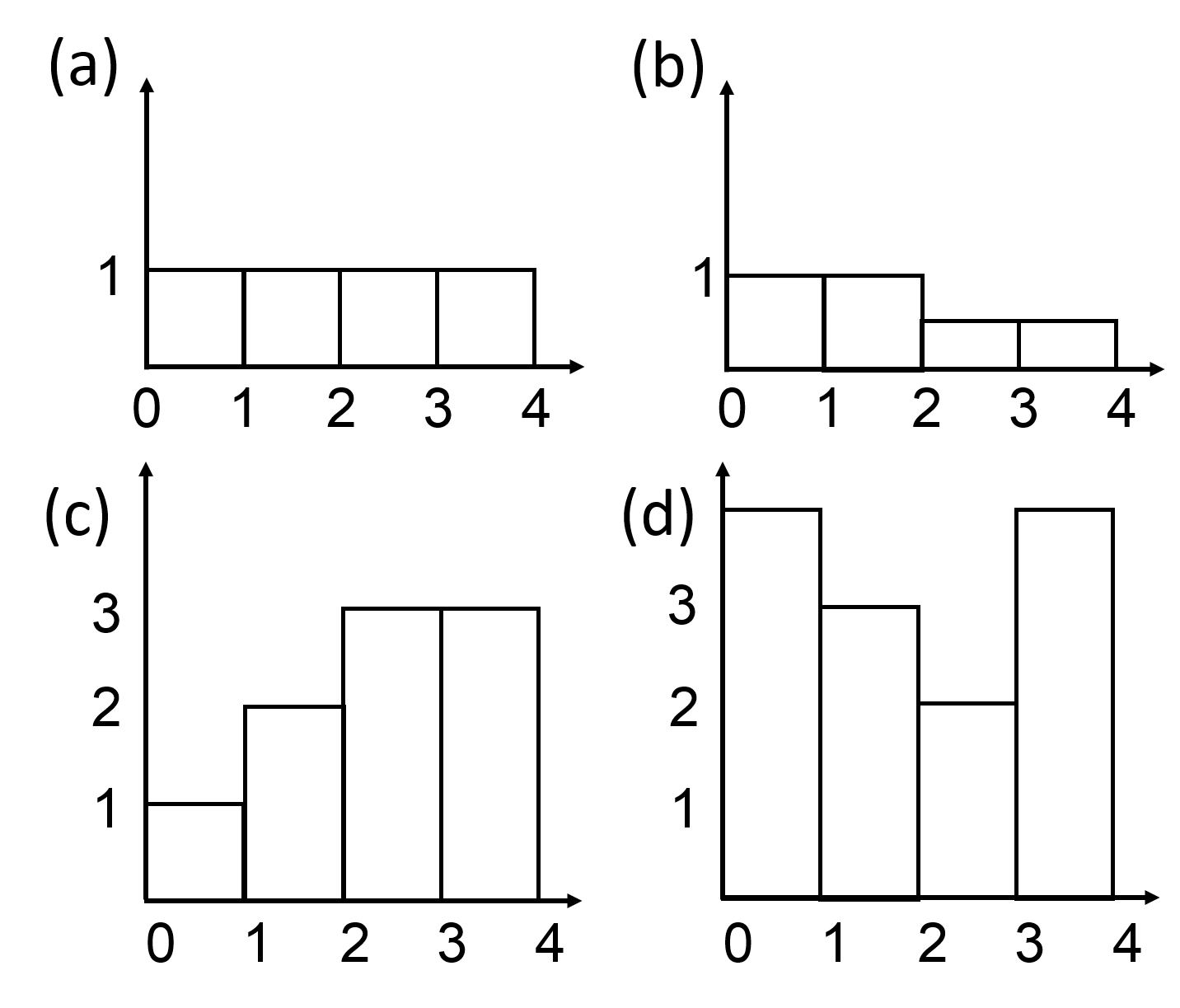}
            \label{fig:ex-1}
            }\\
        \subfigure[Earliest arrivals and latest departures]{
        	\includegraphics[width=0.8\columnwidth]{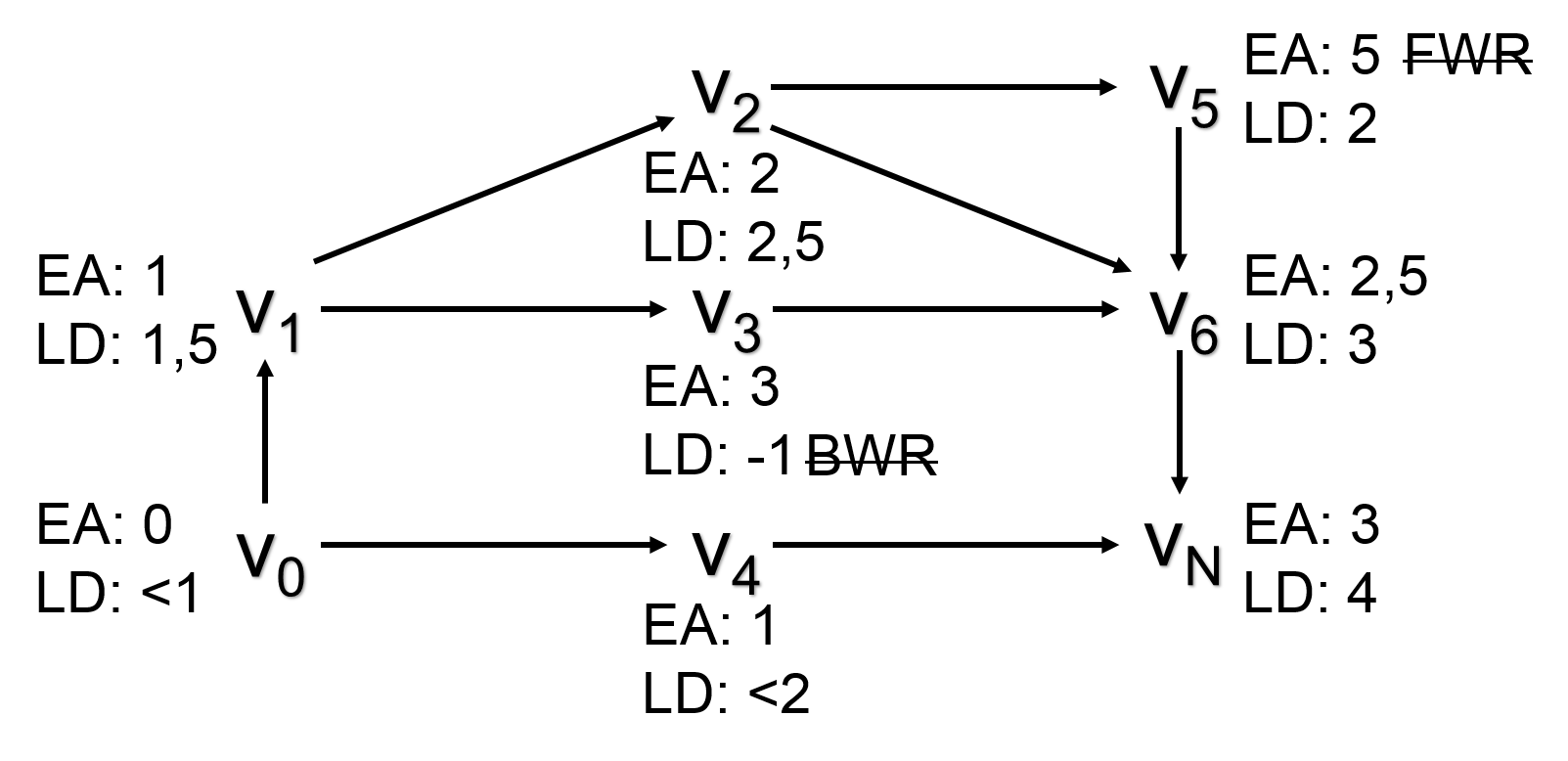}
        	\label{fig:ex-2}
        }
\caption{Example graph for a query with source $v_0$, destination $v_N$,
departure time $0$ and time budget $b = 4$.}
 \label{fig:ex}
\end{figure}

The definitions are visualized in Figure~\ref{fig:ex}. Figure~\ref{fig:ex-0} shows a graph with source $v_0$ and destination $v_N$.
Departure time is $t_0 = 0$, and let the time budget $b = 4$. The time-dependent travel times along the arcs are visualized in Figure~\ref{fig:ex-1}. The arcs
in Figure~\ref{fig:ex-0} are labeled with letters which correspond to the travel time functions in Figure~\ref{fig:ex-1}. Figure~\ref{fig:ex-2} shows the
earliest arrivals and latest departures according to this query. If not indicated otherwise, vertices are forward-reachable as well as backward-reachable. The
earliest arrivals are computed as in the definition. For example, $\ea_1 = 0 + \trv((v_0,v_1),0) = 1$, implying that the fastest path from $v_0$ at $t_0 = 0$
arrives at $v_1$ at time $1$.

%When starting at the query vertice $v_0$, 

Consider the following path:
\[((v_0,v_1),(v_1,v_2),(v_2,v_5),0) = 1 + 1 + 3 = 5 > b\]
Since there is no shorter path from $v_0$ to $v_5$, $v_5$ is not $\fwr{}$ within the given budget. From $v_0$ there are two feasible paths to $v_N$:
\[
((v_0,v_4),(v_4,v_N),0) = 1 + 2 = 3
\]
\[
((v_0,v_1),(v_1,v_2),(v_2,v_6),(v_6,v_N),0) = 1 + 1 + 0.5 + 1 = 3.5
\]
Which of the two constitutes the optimal path depends on the time-dependent value functions and will be discussed later on. 

Let us turn to the computation of the latest departures. For example, $\ld_6 = 0 + 4 - 1 = 3$, implying that in order to arrive at $v_N$ no later than $t_0 +
b$, one must depart at time $3$. Now, consider $\ld_4$: the only path to the destination follows the arc $(v_4,v_N)$. Departing at time $3$ results in exceeding
the budget, since $\trv_{4,N}(3) = 3$. The same holds for time $2$. However, when departing at $t<2$, then $\trv_{4,N}(t)\leq 2$, resulting in an arrival
within the budget $b$. Hence, in order to compute the latest departure time of a vertex $v_i$ a linear scan among the time windows might be required. Note that
the difference between ``at time $2$'' and ``before time $2$'' is infinitesimal. In theory we set the latest departure at $v_4$ to $2-\varepsilon$, in practice, we
adjust it by a sufficiently small number. Let us note that this special case only occurs if the latest departure at a vertex by chance
coincides with the border between two time windows $\tau_k$ which is very unlikely in practice. Finally, we prove that in order to compute all feasible paths
it suffices to consider the vertices which are forward-reachable and backward-reachable.

\begin{lemma}\label{lem:feasible-vertices}
All vertices along all feasible paths are contained in the intersection of the sets of forward-reachable and backward-reachable vertices. 
\begin{proof}
Suppose there exists a feasible path $p$ visiting a vertex $v_i \notin \fwr{} \cap \bwr{}$. Case 1:
$v_i \notin \fwr{}$, this implies $\ea_i > t_0 + b$, meaning that the subpath from $v_0$ to $v_i$
already exceeds the budget, which contradicts the assumption that $p$ is feasible. Case 2: Analogously, if $v_i
\notin \bwr{}$, then $\ld_i < t_0$, hence $p$ cannot be a feasible path.
\end{proof}
\end{lemma}

\begin{algorithm}[t]
    \caption{FWR}
    \label{alg:fwr}
    \begin{algorithmic}[1]
        \Procedure{FWR}{$G, v_0, t_0, t_0 + b$}
        	\State $Q \leftarrow $ empty queue of vertices $v_i$ sorted asc. by $\ea_i$
        	\State result $ = \emptyset$
        	\State $\ea_0 = t_0$
        	\State $Q$.add($v_0$)
        	\While {Q is not empty}
        		\State $v_i \leftarrow $ Q.removeFirst
        		\For {$(v_i,v_j) \in G.A$}
        			\State $\hat{\ea}_j \leftarrow \ea_i + \trv_{i,j}(ea_i)$
        			\If {$\hat{\ea}_j < t_0 + b$}
	        			\If {$\lnot$ result.contains($v_j$)}
	        				\State $\ea_j \leftarrow \hat{\ea}_j$
	        				\State $Q$.add($v_j$)
	        				\State result.add($v_j$)
	        			\Else
	        				\If {$\hat{\ea}_j < \ea_j$}
	        					\State $\ea_j \leftarrow \hat{\ea}_j$
	        				\EndIf
	        			\EndIf
	        		\EndIf
				\EndFor
			\EndWhile
           	\State return result
        \EndProcedure
    \end{algorithmic}
\end{algorithm}

\section{Computing Reachability}
\label{sec:algorithm-1}

Having clarified the importance of the sets of reachable vertices, we will now go into detail on their
computation. Both sets are generated by expansions around source and destination with ``time-radius $b$''. By the above lemma, all
vertices along all feasible paths are contained in the intersection of the two reachability regions.

In order to compute the set of forward-reachable vertices (for a given source $v_0$, departure time
$t_0$, and a time budget $b$), we perform a modified time-dependent all-target Dijkstra search
starting at $v_0$ at time $t_0$,  called \fwr{} and illustrated in Algorithm~\ref{alg:fwr}. All earliest arrival times are initialized as $\infty$ except for
the source $v_0$ for which it is $0$. Additional to a result set containing all visited vertices, we maintain a priority queue containing vertices
sorted in ascending order by the earliest arrival time of each vertex. In every iteration, the top element of the queue is removed. In the \texttt{for}-loop
spanning lines 8-21, the outgoing arcs of the current vertex $v_i$ are explored. In line 9, the travel time
from $v_i$ to its neighbor $v_j$ is added to the earliest arrival time at $v_i$. If this value does
not exceed the budget $b$, $v_j$ is forward-reachable. If $v_j$ was previously not visited, $ea_j$ is
set accordingly and $v_j$ is added to the queue as well as the result set. If $v_j$ was previously visited
and the updated earliest arrival time is better than the previous, it is updated accordingly. Note that once a vertex $v_i$ has been dequeued, the label $ea_i$
is final. This is equivalent to the Dijkstra property which ensures that any processed vertex is reached by the shortest path, in this case the fastest
time-dependent path. Finally, the result set contains all vertices reachable from the source within the budget.

\begin{algorithm}[t]
    \caption{BWR}
    \label{alg:bwr}
    \begin{algorithmic}[1]
        \Procedure{BWR}{$G, v_N, t_0, t_0 + b$}
        	\State $Q \leftarrow $ empty queue of vertices $v_i$ sorted desc. by $\ld_i$
        	\State result $ = \emptyset$
        	\State $\ld_N = t_0 + b$
        	\State $Q$.add($v_N$)
        	\While {Q is not empty}
        		\State $v_j \leftarrow $ Q.removeFirst
        		\For {$(v_i,v_j) \in G.A$}
        			\State $\tau_k \leftarrow \arg\max_{\tau_l} \{\tau_l < \ld_j\}$
        			\While {$\tau_k + \trv_{i,j}(\tau_k) > \ld_j$}
        				\State $k \leftarrow k-1$
        			\EndWhile
        			\State $\hat{\ld}_i \leftarrow \ld_j - \trv_{i,j}(\tau_k) \ [ - \varepsilon]$
        			\If {$\hat{\ld_i} \geq t_0$}
	        			\If {$\lnot$ result.contains($v_i$)}
	        				\State $\ld_i \leftarrow \hat{\ld}_i$
	        				\State $Q$.add($v_i$)
	        				\State result.add($v_i$)
	        			\Else
	        				\If {$\hat{\ld}_i > \ld_i$}
	        					\State $\ld_i \leftarrow \hat{\ld}_i$
	        				\EndIf
	        				
	        			\EndIf
	        		\EndIf
				\EndFor
			\EndWhile
           	\State return result
        \EndProcedure
    \end{algorithmic}
\end{algorithm}

The set of backward-reachable vertices is computed in a similar fashion but slightly more complicated. The procedure is presented in
Algorithm~\ref{alg:bwr} and called \bwr{}. In addition to the result set containing all backward-reachable vertices, we maintain a queue of vertices sorted in
descending order by $\ld_i$. Initially, the latest departure at the destination $v_N$ is set to the latest possible arrival time, $t_0+b$.
The algorithm operates
backwards from the destination, following incoming arcs which are explored in the \texttt{for}-loop
spanning lines 8-25. As was illustrated in the above example (Figure~\ref{fig:ex}), in order to
find the latest departure, it is required to linearly scan the time windows. When
exploring an incoming arc from $v_i$ to $v_j$, $\tau_k$ is initially set to the latest time window not
exceeding the latest departure from $v_j$. If at $\tau_k$ the travel time from $v_i$ to $v_j$ is too
high (i.e., $v_j$ cannot be reached in time for its latest departure), $k$ is decremented implying an earlier departure at $v_i$ but possibly also a different
travel time along the arc.
Note the $[ - \varepsilon]$ in line 13, corresponding to the aforementioned special case that 
$\hat{\ld}_i = \tau_l$ for some $l$. In this case, $\hat{\ld}_i$ is decreased by a small number in
order to avoid computational errors which may occur at time window boundaries. Once a valid latest
departure time for $v_i$ is found, $\ld_i$ is set accordingly. If $v_i$ was previously visited, $\ld_i$ is 
only set if it is later than the previous latest departure. Upon termination, all vertices which are
backward-reachable from the destination within the given budget are in the result set and labeled
with their respective latest departure.

\section{Computing Solutions}
\label{sec:algorithm-2}

\begin{figure}[t]
	\centering
	\subfigure[\forwest{}]{
		\includegraphics[width=0.46\columnwidth]{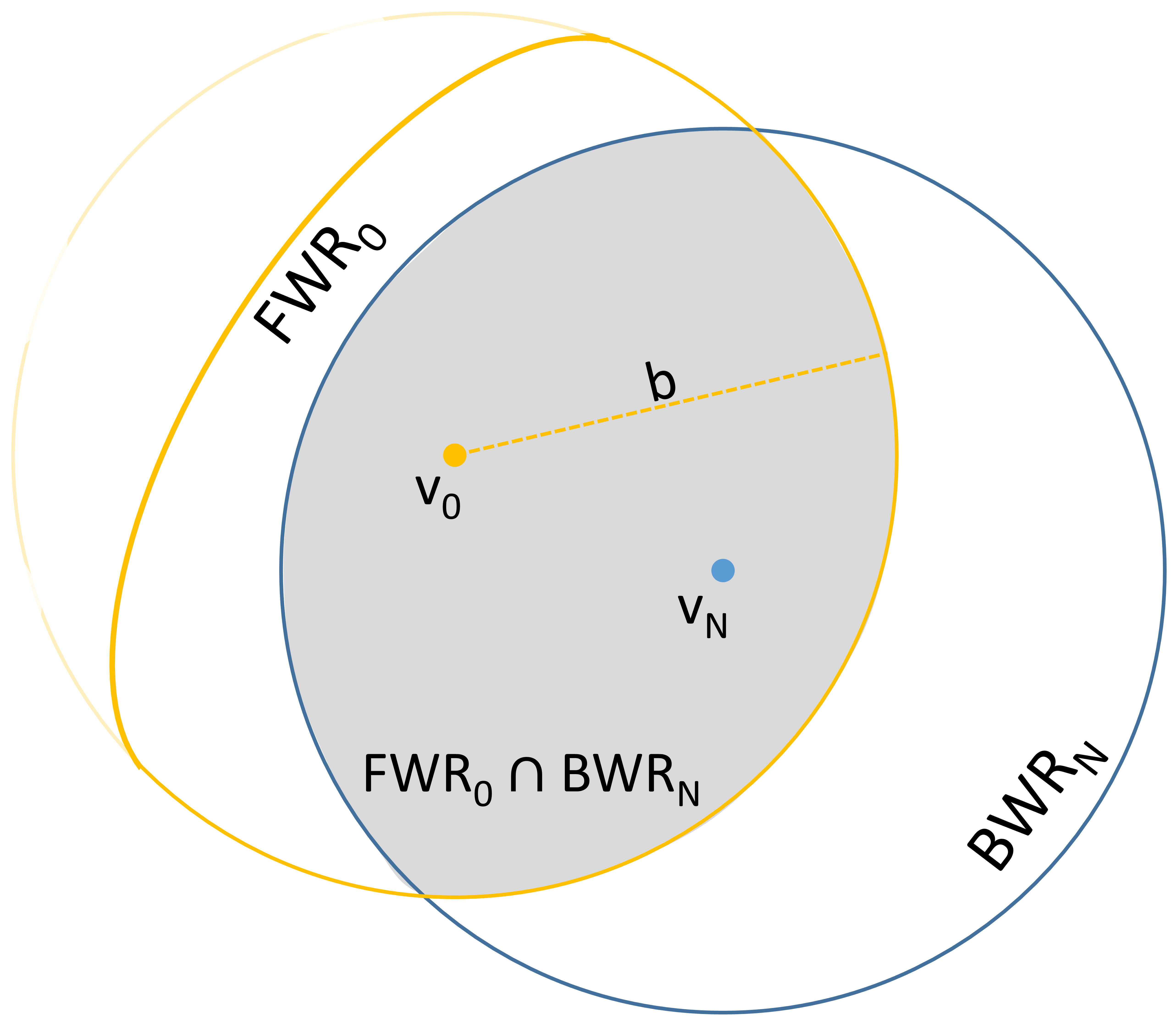}
		\label{fig:small-balls-1}
	}
	\hfill
	\subfigure[\eald{}]{
		\includegraphics[width=0.47\columnwidth]{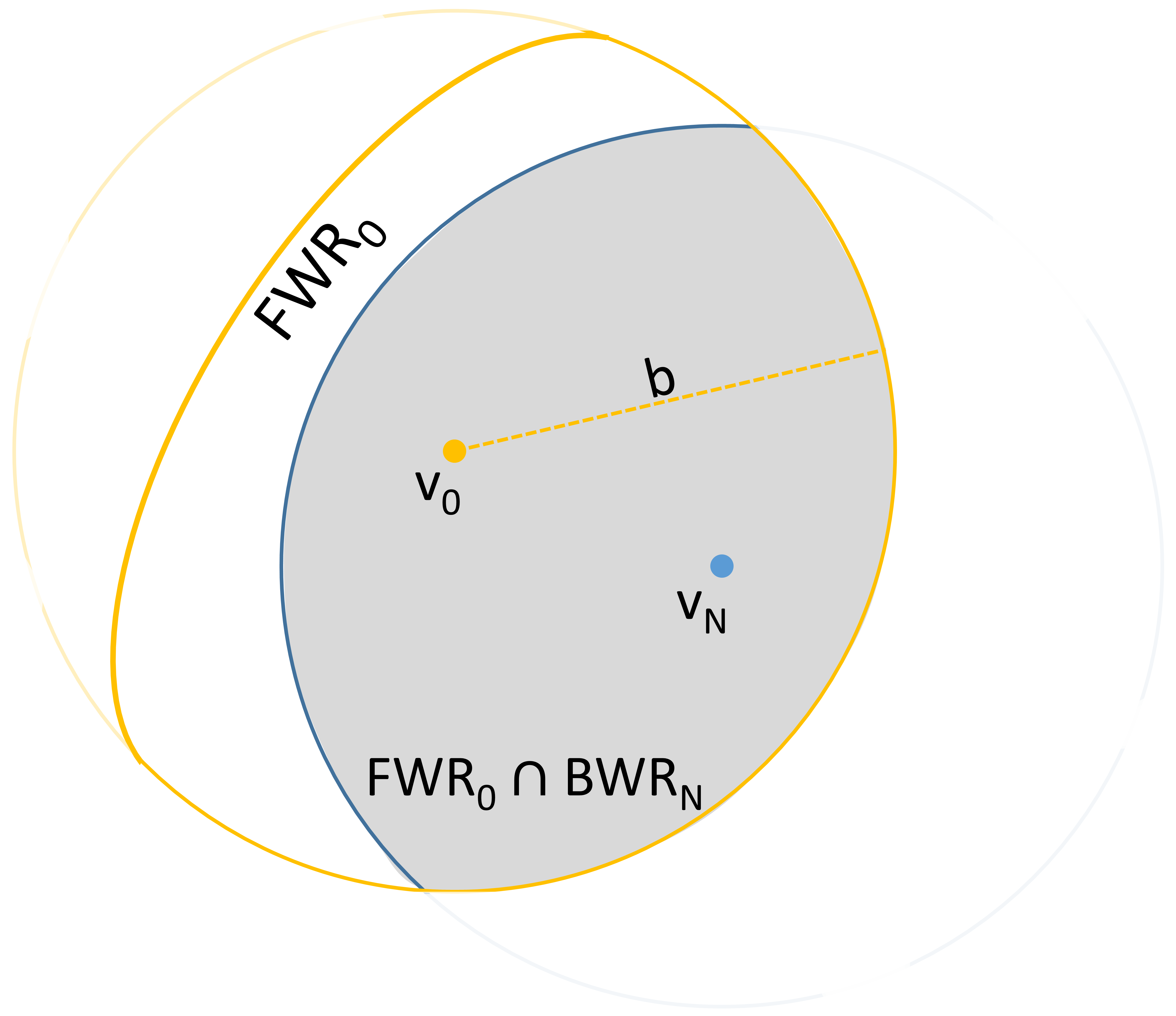}
		\label{fig:small-balls-2}
	}
	\caption{Illustration of the effects of the proposed pruning techniques.}
	\label{fig:small-balls}
\end{figure}

Due to the NP-hardness of the problem, we forgo giving an optimal solution as it would not meet our requirements of providing solutions on large graphs
in efficient time. Instead, we propose a heuristic solution to the \prob{}. So far no other solution to the \prob{} exists. 
Our dynamic programming solution recursively fills gaps with arcs in order to improve the 
value while not exceeding the budget. Upon initialization, the set of 
arcs \arcs{} is empty, the set of gaps \gaps{} contains $(v_0,v_N)$, i.e., 
the gap given by the query. Assume that in the first iteration the arc $(v_i,v_j)$
is inserted. Then \arcs{} holds the new arc $(v_i,v_j)$ and \gaps{} holds
two new gaps, $(v_0,v_i)$ and $(v_j,v_N)$. Into the new gaps, further arcs will
be inserted. This procedure is repeated until the budget is exhausted. Hence,
the \prob{} is solved in a divide-and-conquer manner by recursive insertions.
We therefore refer to the algorithm, pseudo-coded in Algorithm~\ref{alg:main}, as \algo{}.

\begin{algorithm}[t]
	\caption{\algo}
	\label{alg:main}
	\begin{algorithmic}[1]
		\Procedure{\algo}{$G$, \gaps{}, \arcs{}, $b$}
		%\State $G' \leftarrow $ null
		\State bestArc $\leftarrow ($null, null$)$
		\State bestGap $\leftarrow ($null, null$)$
		\State bestRatio $\leftarrow 0$
		%\State detour $\leftarrow 0$
		\For{$(v_i, v_j) \in \gaps{}$}
			\State $\fwr_i \leftarrow \fwr(G, v_i, b)$
			\State $\bwr_j \leftarrow \bwr(G, v_j, b)$
			%\State //Intersection can be integrated into computation of \bwrj
			\State $G_{ij}' \leftarrow \fwr_i \cap \bwr_j$
			\For{$(v_m, v_n)$ in $G_{ij}'$: $\ea_m + \trv_{m,n}(\ea_m) < \ld_n$}
				%\For{$(v_m, v_n) \in \{G_{ij}'.A \mid \ea_m + \trv_{m,n}(\ea_m) < \ld_n\}$}
				\State $b_- \leftarrow \ld_n - \ea_m - \trv_{m,n}(\ea_m)$ 
				\State $val_+ \leftarrow \val_{m,n}(\ea_m)$ 
				\State ratio $ \leftarrow val_+ \cdot b_-^{-1} $
				\If {ratio $>$ bestRatio}
					\State bestRatio $ = $ ratio
					\State bestArc $ = (v_m, v_n)$
					\State bestGap $ = (v_i, v_j)$
				\EndIf
			\EndFor
		\EndFor
		\State gaps.remove(bestGap)
		\If{$bestArc = null$}
			\State compute fastest path to close all gaps
			\State \textbf{return} concatenation of fastest paths
		\Else
			\State gaps.add($(v_i,v_m)$)
			%\Comment{\parbox[t]{.5\linewidth}{add $($bestGap$_1,$bestArc$_1)$ at former position of $(v_i,v_j)$}}
			\Comment{\parbox[t]{.5\linewidth}{at former index of bestGap}}
			\State gaps.add($(v_n,v_j)$)
			\Comment{\parbox[t]{.5\linewidth}{after $(v_i,v_m)$}}
			\State arcs.add($(v_i,v_j)$)
			%\Comment{\parbox[t]{.5\linewidth}{bestArc}}
		\EndIf
		\State $\algo(G,\gaps{}, \arcs{}, b_-)$
		\EndProcedure
	\end{algorithmic}
\end{algorithm}

We will now explain in detail how the algorithm proceeds. The two sorted lists
\gaps{} and \arcs{} hold two-tuples of vertices. \gaps{} holds pairs between
which arcs are to be inserted, and \arcs{} holds the converse information, i.e.,
the arcs already inserted. With every insertion, the budget is decreased by the
duration of the detour that has to be taken to insert the new arc. However, this 
detour is not explicitly computed as a path. Instead, it is implicitly derived
from the information generated during \fwr{} and \bwr{}. The earliest arrivals
and latest departures of every gap allow to decide whether an arc can be \emph{feasibly
inserted}, i.e., without exceeding the budget. This is crucial to the performance
of our algorithm. Established metaheuristics usually compute, improve and perturb
solutions. Starting from an  initial path, promising arcs are (usually randomly)
inserted and under certain conditions (usually randomly) deleted, to avoid local
value maxima. Instead,  \algo{} holds a sorted list of arcs which may be feasible
inserted into a path. The actual paths between these arcs, however, are
not computed until the budget is exhausted.

The arc which is to be inserted into a gap is chosen based on a heuristic. Among all arcs
which can be feasibly inserted, the arc is selected which yields the best ratio of 
value to detour. This step is performed in the \texttt{for}-loop spanning lines 9 to 19.
As established, all feasibly insertable arcs lie in the intersection of the reachability regions.
The intersection is performed in line 8, therefore the set in line 9 comprises all arcs
which can be feasibly inserted. In line 10, the detour incurred by inserting an arc is
expressed as a decrease in budget. Assume that for the gap between $v_i$ and $v_j$,
the arc $(v_m,v_n)$ is examined for insertion. $ea_m$ is the earliest arrival
at $v_m$ coming from $v_i$, and $ld_n$ is the latest departure from $v_n$ to reach $v_j$ 
``in time'' (depending on the recursive budget $b$). Thus, $ld_n - ea_m - \trv_{m,n}(ea_m)$
corresponds to the budget that would remain when taking the detour of traveling from 
$v_i$ to $v_m$, then along $(m,n)$ (at time $ea_{m}$) and from $v_n$ to $v_j$.
Hence, if inserting $(v_m,v_n)$, the budget will be set to $ld_n - ea_m - \trv_{m,n}(ea_m)$.
In line 11, the gain in value is estimated. Note that the actual gain in value may
be higher (due to other arcs along the path). However, in order to compute the actual
gain in value, the fastest paths to and from each arc would have to be computed. Also, the value
of all succeeding arcs would have to be reassessed. For the 
sake of efficiency, we therefore only estimate the value as proposed. The estimated gain in 
value divided by the decrease in budget yields the ratio by which we evaluate the arcs. 
The arc with the best ratio is inserted, i.e., the arc is added to \arcs{} as a pair of vertices, 
and two pairs of vertices are added to \gaps{} (the gaps left and right of the new arc). 
As long as at least one arc qualifies for insertion, \algo{} is recursively called
with the extended sets \gaps{} and \arcs{} and the reduced budget $b_-$. If no arc can be
inserted (into any of the gaps), the time-dependent fastest paths for all gaps are
computed and the concatenation is returned which constitutes our approximate solution
to the \prob{}.

\subsection{Pruning Strategies}

Let us discuss three variations which are omitted in the algorithm description for reasons
of brevity. $(i)$, it is recommended to employ a simple forward estimation during \fwr{}.
$(ii)$, it is recommended to compute the intersection of \fwri{} and \bwrj{} during computation
of \bwrj{}. $(iii)$, it is possible to make the recursive call with a restricted portion of the graph.
Variations $(i)$ and $(ii)$ are based on the following observation. The sets \fwri{} and \bwrj{}
are not needed separately, only their intersection is needed. Therefore, we may exclude 
vertices which are not contained in the intersection during computation of the respective sets.

Regarding $(i)$: We propose to employ a simple forward estimation for pruning. If 
information about the speed limits is available, the following lower bound can 
be used with negligible overhead.
During \fwr{}, in line 10, the \texttt{if}-clause may be tightened:
\[\ea_j + \frac{ed_{jN}}{ms} < t_0 + b\]
where $ed_{jN}$ is the Euclidean distance from $v_j$ to the destination $v_N$ 
and $ms$ is maximum speed in the given network. If traveling from $v_j$ to the 
destination at maximum speed exceeds the budget, so will any actual path. Any 
such vertex cannot be forward-reachable. Figure~\ref{fig:small-balls-1} illustrates 
the effect of this pruning strategy yielding a trimmed forward-reachability region. We refer to this pruning strategy as \emph{\forwest{}} pruning.
At marginal computational cost, \forwest{} yields a perceptible reduction in candidates. In our experiments, over 15\% of vertices are pruned during \fwr{}
using this approach. Of course, the method holds analogously for \bwr{} and can equally be applied. However, during \bwr{} it is recommended to use the
following pruning technique which is superior but exclusive to \bwr{}.

Regarding $(ii)$: One may use that for all vertices in the intersection 
$\fwr_i\cap\bwr_j$ holds $\ea_i\leq \ld_i$.
This follows from the definition:
\begin{eqnarray*}
&\ea_i > \ld_i\\
&\Leftrightarrow t_0 + \trv(p^\star_{0i}) > t_0 + b - \trv(p^\star_{iN})\\
&\Leftrightarrow \trv(p^\star_{0i}) + \trv(p^\star_{iN}) > b
\end{eqnarray*}
where $p^\star_{ij}$ denotes the fastest path from $v_i$ to $v_j$ at $\ea_i$. Hence, the 
\texttt{if}-clause in line 14 of \bwr{} can be tightened to $\ld_i \geq \ea_i$. In 
this case, the simple forward estimation cannot be applied because 
\[\ea_i \geq t_0 \frac{ed_{0i}}{ms} \geq t_0\]
By construction, $ea_i$ is based on an actual path expansion and therefore serves as 
a tighter bound than any estimation. We refer to this pruning strategy as \eald{} pruning. Figure~\ref{fig:small-balls-2} illustrates the effect
of this pruning strategy, where the vertices which are not forward-reachable are not visited
during \bwr{}. Thus, instead of computing both reachability regions separately and subsequently
their intersection, the information of \fwr{} may be used in \bwr{}. This limits the search space
and implicitly computes the intersection. Every vertex in the intersection will be visited during 
\bwr{}. Unvisited vertices are either not forward-reachable or backward-reachable. The effect of this pruning strategy is remarkable. Particularly when taking
into account that it causes no computational overhead. When visiting a vertex during \bwr{}, it suffices to retrieve its earliest arrival to decide if it
qualifiies for pruning.

Regarding $(iii)$: It is possible but not necessarily recommended to issue the recursive
call of \algo{} with a restricted portion of the graph. As established in 
Lemma~\ref{lem:feasible-vertices}, all vertices along feasible paths are in the
intersection of the reachability regions. This property holds recursively. For instance,
consider the insertion of a first arc $(v_i,v_j)$ into the initial gap $(v_0,v_N)$. In
the next recursion, two gaps will be examined, $(v_0,v_i)$ and $(v_j,v_N)$, as illustrated
in Figure~\ref{fig:balls-2}. Consider the
first gap $(v_0,v_i)$, the respective reachability regions (forward from $v_0$ and 
backward from $v_i$) are contained in the original reachability regions (forward from
$v_0$ and backward from $v_N$. Moreover, the intersection is contained in the original 
intersection. Thus, it suffices to input a restricted graph into the next recursion:
\[G_{\text{next}} = \bigcup \{G_{mn}' \mid (v_m,v_n)\in\gaps{}\}\]
Albeit correct, actually computing the union of these graphs may cause more 
computational overhead than it saves. In our experiments, computing the restricted 
graph has not proved efficient. We input the initial intersection of $\fwr_0$ and
$\bwr_N$ into all recursions as it restricts the graph significantly while not requiring
costly unions of graphs.

Let us note that Gunawan et. al use very similar 
terminology in their approach to solving the TD-OP \cite{TDOP-MathematicalModelMetaheuristicsTDOP-Gunawan2014}.
However, there are significant differences to the approach presented here. First,
the authors solve TD-OP where travel times are time-dependent but values are 
assumed to be static and associated with vertices. Second, the authors propose a different 
algorithmic approach. While also using the terms earliest arrival and latest departure,
they are computed differently and for different purposes. Given a specific path,
a routine called \textsc{ForwardPropagation} computes the earliest arrival
times at all vertices along the path. Analogously, \textsc{BackwardPropagation} computes the latest
departures. In comparison, our \fwr{} and \bwr{} procedures compute earliest arrivals and 
latest departures for all vertices which are feasibly insertable into a given gap. From this
information, the travel time for a detour to any feasibly insertable arc
is derived. This is done without explicit path computation.
While we use dynamic programming in a divide-and-conquer manner, Gunawan et. al follow
an Iterative Local Search (optionally Variable Neighbor Descent) approach. where actual
paths are computed, improved and perturbed. In their experiments, they produce near-optimal
to optimal solutions with a maximum runtime of one second. Their approach is evaluated
on small data sets with at most 40 vertices and 40 time windows. The arcs correspond
to precomputed fastest paths. This restrictive scenario is not comparable to the use
case we are modeling in this paper where fast response times are required for a more 
complex problem on graphs which are orders of magnitude larger.

\begin{figure}[t]
	\centering
	\includegraphics[width=0.98\columnwidth]{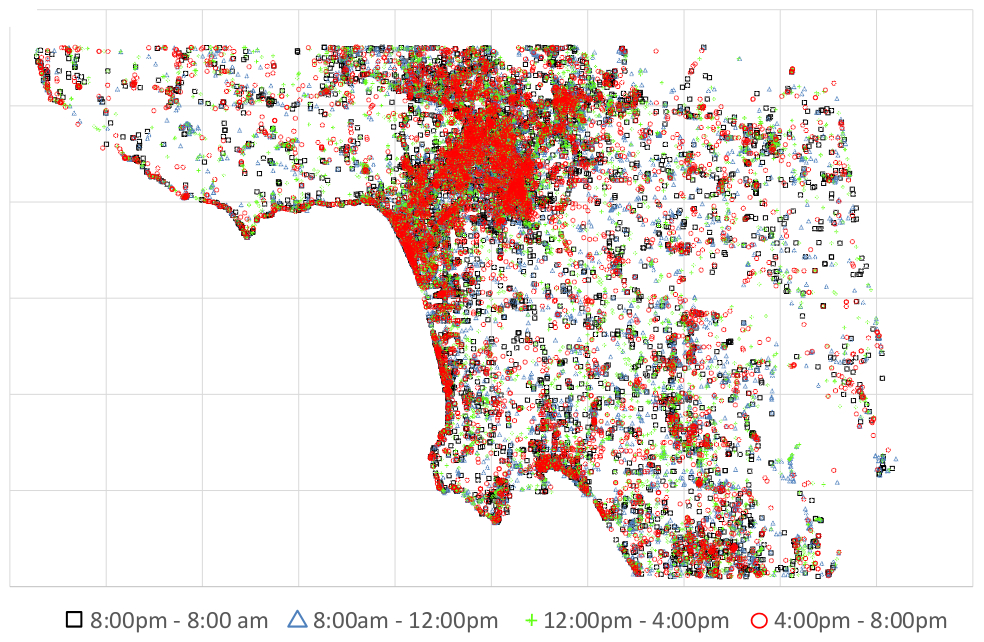}
	\caption{Value arcs in experimental road network of Los Angeles. Each value arc corresponds to a colored
		shape which indicates the arc's value peak interval.}
	\label{fig:arcs}
\end{figure}

\section{Experimental Evaluation}
\label{sec:exp}

We evaluate our solution to the \prob{} on a time-dependent road network of Los Angeles, CA, USA
\cite{TD-CaseForTimeDepRoadNetworks-FoundationPaper-DEMIR-SHAHABI-GIS10,TD-BigDataAndItsTechnicalChallenges-SHAHABI-14} which contains 
about 500K vertices and 1M arcs. The travel time functions in this network are derived from large-scale and
high-resolution (both spatial and temporal) sensor data (both live and historic) from different transportation authorities
in California. The step length of the piecewise constant travel time functions is 5 minutes, details can be found in 
\cite{TD-OnlineComputationOfFastestPathsInTimeDependentSpatialNetworks-DEMIRYUREK-11}. Based on the data, streets are uncongested between 9 pm and 6 am.
Therefore, travel times are assumed to be static in this interval.

\begin{table}[t]
	\begin{center}
		\begin{tabular}{ l | c | c }
			\toprule
			peak interval & \# of arcs & \% of all arcs \\%\% of all non-zero value arcs & 
			8:00 am - 12:00 pm & 6,238 & 0.57\%\\%& 24.1\% 
			12:00 pm - 4:00 pm & 7,366  & 0.67\%\\%& 28.5\%
			4:00 pm - 8:00 pm & 5,851 & 0.53\%\\%& 22.6\%
			8:00 pm - 8:00 am & 6,417 & 0.59\%\\%& 24.8\% 
			all intervals & 25,872 & 2.38\%\\
			\bottomrule
		\end{tabular}
	\end{center}
	\caption{Statistics about arcs with non-zero value.}
	\label{tab:arcs}
\end{table}

\begin{figure}[t]
	\centering
	%\raggedleft
	\includegraphics[width=0.9\columnwidth]{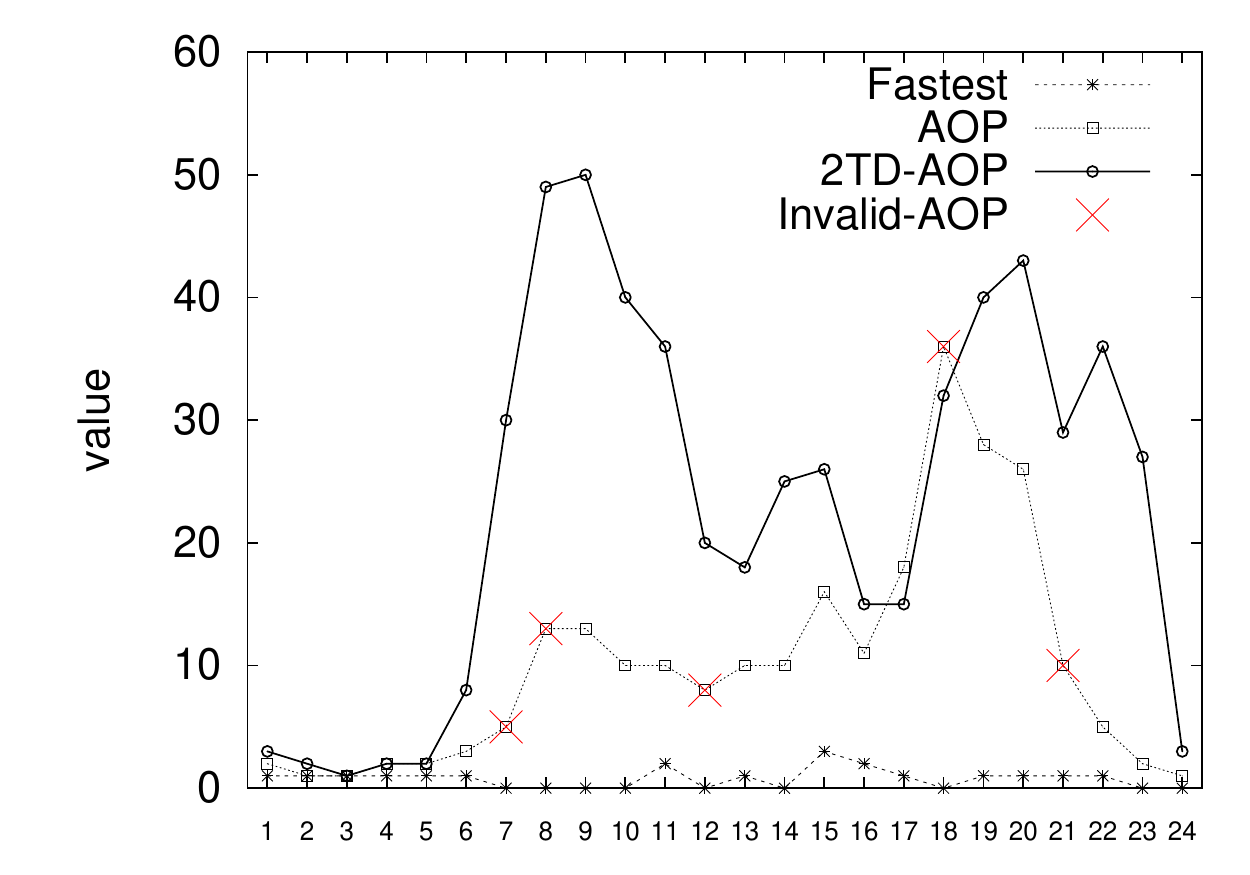}
	\caption{Value results for the introductory example in Figure~\ref{fig:motiv}.}
	\label{fig:motiv-values}
\end{figure}

The time-dependent value functions
are derived from a geo-tagged set of Flickr photos \cite{FlickrPhoto:MMSys2014} consisting of 217,391 photos in the area of Los Angeles. For each arc
$(v_i,v_j)$ the value $\val_{i,j}(t)$ is the number of photos which have $(v_i,v_j)$ as their nearest arc (but not
further than one kilometer) and which were taken in the hour-interval that contains $t$. Hence, the step length
of the piecewise constant value functions is 60 minutes. Figure~\ref{fig:arcs} illustrates the size of the network
and the distribution of non-zero value arcs. Each arc which has positive value in some time window is visualized as
a colored shape. Shape and color depend on the time interval in which the value of the arc is the highest. For example, an arc with
highest value between 6:00 and 7:00 pm will be displayed as red circle. Table~\ref{tab:arcs} presents the absolute and
relative numbers of arcs with non-zero value. The number of photos taken during the day is about three times higher than during night.

For the following experiments, we arrange paths into six time buckets. We randomly draw source and destination vertices from the network and compute the
respective time-dependent fastest paths using \cite{TD-OnlineComputationOfFastestPathsInTimeDependentSpatialNetworks-DEMIRYUREK-11}. If the travel time of a
path is $k$ minutes $\pm\varepsilon$, its source and destination pair is assigned to the $k$ minutes time bucket for $k\in\{5,10,15,20,25,30\}$. Each time
bucket consists of twenty source and destination pairs. Since travel times are empirically static at night and value peaks less frequent at night, we randomly
sample departure times from the interval 8 am to 8 pm. We make a case for twofold time-dependence, therefore, a network with significant time-dependent impact
on travel time and value is essential. Our standard experimental setting has the following inputs. Standard time bucket is 20 minutes, standard query budget is
200\% (i.e., 40 minutes), and departure times are randomly drawn from 8 am to 8 pm. Deviating values are explicitly mentioned. For each query, the departure
time is the same across time-dependent algorithms.

We omit providing results of the MIP formulation given in Section~\ref{sec:problem-definition} generated by a complex solver. This is due to the exorbitant
computatial requirements. Even for small instances with at most 100 vertices, solving a less complex MIP often requires days
\cite{TDOP-YING-AntColony-DriveSpeedModel-FastSolutionToTDOP-OR14-VANSTEENWEGEN,TDOP-MathematicalModelMetaheuristicsTDOP-Gunawan2014}. However, for comparison,
we compute optimal solutions with a custom dynamic programming method.

\begin{figure}[t]
	\centering
	\includegraphics[width=0.98\columnwidth]{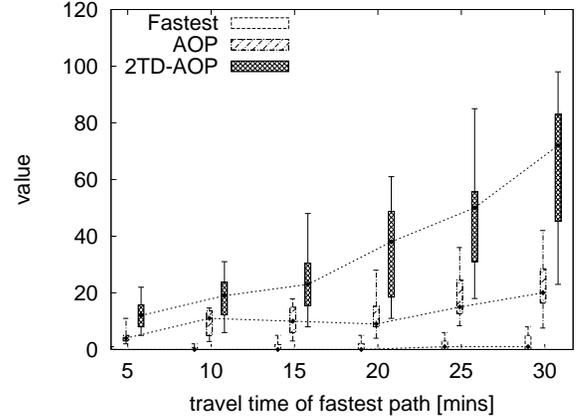}
	\caption{Value results for varying fastest path lengths.}
	\label{fig:value}
\end{figure}

\subsection{Experimental Results}
\label{subsec:results}

In a first set of experiments, we compare the value of result paths generated by different algorithms taking varying degrees of information into account. As in
the introductory example, we compare \prob{} paths computed by \algo{} with static scenic paths and to time-dependent fastest paths.
For the static scenic paths, we use the AOP solution presented in \cite{AOP-Ying-Shahabi-ACMGIS-15} which employs an Iterative Local Search approach combined
with spatial pruning techniques. In accordance with the metaheuristic, it generates an initial solution which is subsequently improved by inserted arcs and
perturbed by deleting arcs. The insertion follows a heuristic, the deletion is pseudo-random. We refer to solutions created by this approach as \static{} paths
and call this algorithm AOP-ILS. It operates on a network with static travel time and static value which are both derived by averaging the respective
time-dependent functions. Note that \static{} queries are given the same time budget but require no departure time.
For the time-dependent fastest paths, we use the solution proposed in \cite{TD-OnlineComputationOfFastestPathsInTimeDependentSpatialNetworks-DEMIRYUREK-11}
which relies on hierarchical routing and forward estimations to conduct efficient bidirectional path computation. We refer to the results as \fastest{} paths.
The results generated by \algo{} are referred to as \twotd{} paths.

The values and travel times of all paths are computed w.r.t.\ the time-dependent functions. For a given path, the value and travel time are the
sum of value and travel time of each arc along the path upon arrival. Recall the introductory example in Figure~\ref{fig:motiv}. For varying departure times in
this example, the result values are displayed in Figure~\ref{fig:motiv-values}. As mentioned in the example, it is possible that paths computed in the static
network are invalid in the time-dependent network. This is the case if the time-dependent travel time exceeds the budget. Static paths where this is the
case are marked with a red cross. For instance, when departing at 6 pm, the time-dependent travel time is 37 minutes, exceeding the budget of 25 minutes by
almost 50\%.

\begin{figure}[t]
	\centering
	\includegraphics[width=0.98\columnwidth]{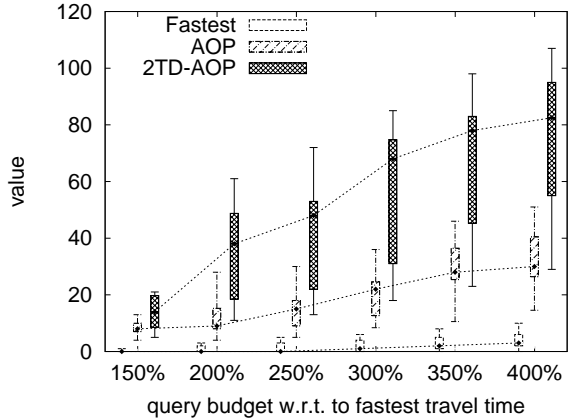}
	\caption{Value results for varying time budget.}
	\label{fig:budget}
\end{figure}

Figure~\ref{fig:value} shows the value for the different approaches. \fastest{} paths serve a baseline which shows how much value
can be attained ``by chance'' as it does not optimize for value. Indeed, \fastest{} paths collect negligible value. Since the non-zero value arcs are rather
scarce, this is not surprising. \twotd{} and \static{}, in contrast, optimize for value. It can be observed that on average \twotd{} generates three times the
value of \static{}. This gap widens for increasing path length.

Figure~\ref{fig:budget} illuminates the effect of the query budget. We observe that with increasing budget, \twotd{} paths generate significantly more value
than \static{} solutions. This is particularly noteworthy as the problem becomes more complex with increasing budget. A higher budget allows for greater
detours which in turn increases the search space. \algo{} leverages this effect to create better solutions. In contrast, \static{} results barely improve.
Therefore, albeit increasing complexity, twofold time-dependence is able to enhance results considerably.

As mentioned before, static paths may prove invalid when considering time-dependent travel time. Given the departure times of \twotd{} and \fastest{}, we
determine the portion of static paths that prove invalid when evaluated in the time-dependent network. The percentage of \static{} which are invalid increases
from 25\% to almost 50\% across the increasing time buckets. Note that Figures~\ref{fig:value} and \ref{fig:budget} show the values of all \static{} paths,
both valid and invalid, as there is no significant difference. However, taking into account the percentage of invalid results and the low overall value, static
results cannot compete with twofold time-dependent results.

\begin{figure}[t]
	\centering
	%\raggedleft
	\includegraphics[width=0.9\columnwidth]{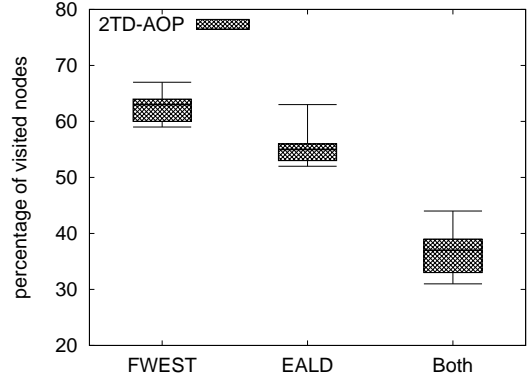}
	\caption{Percentage of visited vertices when employing the proposed pruning techniques.}\label{fig:pruning}
\end{figure}

The proposed pruning strategies are evaluated in Figure~\ref{fig:pruning}. It shows the percentage of visited vertices relative to the number of visited vertices
when employing no pruning technique. In this case, all vertices in \fwr{} and \bwr{} have to be visited. In Section~\ref{sec:algorithm-2} we propose two pruning
techniques for \algo{}, \forwest{} and \eald{}. \forwest{} is a basic forward estimation which can be applied during \fwr{} and \bwr{}. \eald{} uses the
information generated during \fwr{} to limit the search space of \bwr{} and can only be applied during \bwr{}. During \bwr{}, \eald{} is superior to
\forwest{}. Hence, when both are available, \forwest{} is obsolete during \bwr{}. In the experiments displayed in Figure~\ref{fig:pruning}, we show the effect
of each pruning technique separately as well as of both pruning techniques combined. When only employing \forwest{} (during both \fwr{} and \bwr{}), almost 40\%
of the vertices can be pruned. Although \eald{} can only be applied during \bwr{}, it prunes about 45\% of the vertices. As \forwest{} causes negligible and \eald{}
causes no computational overhead, it is recommended to employ both. Combining both approaches, 65\% of all vertices in \fwr{} and \bwr{} can be pruned.

\begin{figure}[t]
	\centering
	\includegraphics[width=0.98\columnwidth]{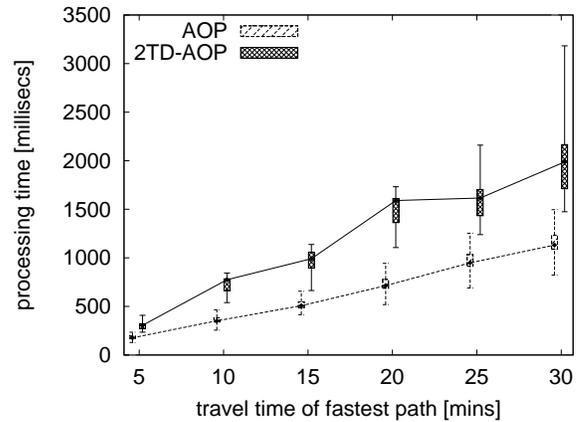}
	\caption{Processing time for varying path lengths.}\label{fig:runtime}
\end{figure}

In terms of general complexity, the \prob{} is considerably more intricate than the AOP. While both problems are NP-hard, twofold time-dependence incurs
additional complexity. This is mainly due to two aspects. First, fastest path computation is more efficient in static networks. Second, in order to assess the
time-dependent value of an arc, it does not suffice to simply retrieve its value. The time frame of arrival has to be considered. In light of these challenges,
the efficiency of \algo{} is all the more surprising. Figure~\ref{fig:runtime} shows the processing times of \algo{} solving the \twotd{} and AOP-ILS solving
the \static{}. \twotd{} solutions can be computed in less than twice the processing time of \static{} solution, usually under two seconds. The increase in
processing time for both algorithms in linear in the travel time of the fastest path. Taking the result quality into consideration, the additional processing
time of \algo{} compensates for the often invalid \static{} paths with less value. It should also be noted that in an application where fixed response times
are required, it would be possible to terminate the arc insertion in \algo{} prematurely and construct an early result path.

\begin{figure}[t]
	\centering
	\includegraphics[width=0.95\columnwidth]{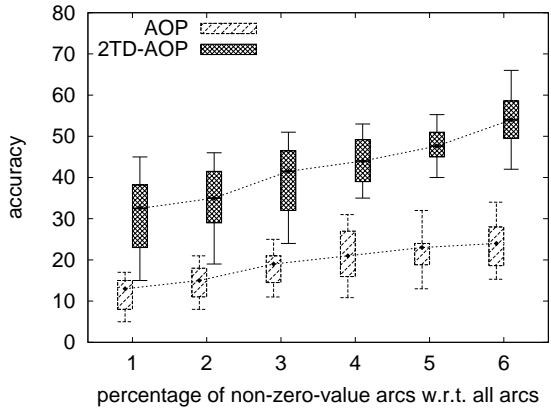}
	\caption{Accuracy of result values relative to optimal results when varying value arc density.}
	\label{fig:accuracy}
\end{figure}

%ACCURACY?
Finally, we compare \prob{} and \static{} paths to optimal solutions in terms of their value. The optimal solutions are computed with a custom
dynamic programming approach using a depth-first search and several pruning techniques. Any na\"ive algorithm, much like a complex solver, would not be able
to compute solutions in feasible time. Our custom approach takes between 3 to 6 hours per optimal path computation for the standard query setting. Even for
minor instances where the fastest path takes 5 minutes and the budget is 10 minutes, computing an optimal solution takes between 1 and 2 hours. Hence,
generating optimal solutions to the \prob{} is infeasible. This is irrespective of the application. Users will not tolerate response times in the hours, and
precomputation is not possible in time-dependent networks. Figure~\ref{fig:accuracy} compares the accuracy of \prob{} paths generated by \algo{} and of
\static{} paths generated by AOP-ILS. Values are shown relative to values of optimal paths. Figure~\ref{fig:accuracy} also illustrates the effect that the
density of non-zero value arcs has on the result. The value is given relative to the optimal value for different networks when varying the percentage of
non-zero value arcs. In our original network about 2\% of the arcs have non-zero value during some time window. For this experiment, we randomly copied value
functions to arcs with zero value to increase density to 4\%, 6\%, 8\% and 10\% of all arcs. Also, we set some of the value functions to zero, generating a
network with 1\% value arcs. As both algorithms follow a heuristic during arc insertion, they are more easily sidetracked in a network with many value arcs.
When the density of value arcs decreases, however, the employed heuristics prove effective. In this case, \algo{} result paths attain between 50\% and 60\%
accuracy. The static solutions, in contrast, hardly exceed 25\%. Again, this establishes the superiority of \prob{} over \static{} solutions. While computing
an optimal solution takes up to 6 hours, \algo{} takes about 2 seconds. Thus, \algo{} produces paths with about 50\% accuracy in $10^{-5}$ the processing time.
As the \prob{} cannot be expected to be solved optimally in efficient time, \algo{} yields a promising trade-off.

In conclusion, our experimental evaluation has established the importance of twofold time-dependence. Solutions to the \prob{} are superior to solutions to the
static AOP. For increasingly complex query settings, this trend is intensified. Furthermore, \algo{} computes solutions that achieve 50\% accuracy compared to
optimal results within seconds. The efficiency can essentially be attributed to the pruning techniques employed in \algo{}. As of now, no time-dependent AOP
has been evaluated on large-scale real-world road networks. We prove that even in such networks, the higly complex \prob{} can be feasibly solved.

\section{Conclusions}
\label{sec:conclusion}

In this paper, we present the Twofold Time-Dependent Arc Orienteering Problem (\prob{}) a novel extension of the family of Orienteering Problems (OP). In this
family of NP-hard graph problems, the goal is to find a path from a given source to a given destination within a given time budget which, additionally,
maximizes the value collected along the way. In comparison to static versions of this problem, \prob{} allows for travel times and value functions to be
time-dependent. The incorporation of time-dependent values is a novel extension which has not been studied yet. The importance of time-dependent values is
showcased and experimentally substantiated. Due to the NP-hardness and the twofold time-dependence, we provide a heuristic approximation for solving the
\prob{}. Our approach is evaluated on a large-scale real-world road network. Thus far, no time-dependent variant of the AOP has been evaluated on a
network of this dimension. Twofold time-dependent results gain significantly more value than static results, showing the importance of twofold time-dependence
empirically. Our algorithm produces result paths with 50\% accuracy where no optimal solution is feasible.

\footnotesize
\vspace{1eM}
\bibliographystyle{abbrv}
\bibliography{literature}

\balance

\end{document}